\documentclass[a4paper,fleqn]{cas-sc}

\usepackage[authoryear,longnamesfirst]{natbib}
\usepackage{amssymb}
\usepackage{amsmath}
\usepackage{tabularx}
\usepackage{cancel}
\usepackage{multirow}
\usepackage{array} 
\usepackage{booktabs}
\usepackage{supertabular}
\usepackage{algorithm}
\usepackage{algorithmic}
\usepackage[shortlabels]{enumitem}
\usepackage{amsthm}
\usepackage{float}
\usepackage{subfig}
\usepackage{graphicx}
\usepackage{rotating}
\usepackage{setspace}
\usepackage{xcolor}
\numberwithin{equation}{section}
\usepackage{ulem}
\usepackage{multirow}

\def\tsc#1{\csdef{#1}{\textsc{\lowercase{#1}}\xspace}}
\tsc{WGM}
\tsc{QE}

\newcommand{\R}{\mathbb{R}}  %% real space

\newcommand{\Var}{\mathbb{V}\mathrm{ar}}  %% variance
\newcommand{\indexSet}{I}   %% index set
\newcommand{\Tr}[1]{\mathrm{Tr}\left[ #1 \right]}

\newcommand{\nvar}{d}   %% number of input variables
\newcommand{\nOutputDimensions}{L}  %% number of 
\newcommand{\indPoints}{i}

\newcommand{\nDoE}{n}  %% number of training set
\newcommand{\SobolNormalized}[1]{S_{#1}}  %% normalized Sobol
\newcommand{\SobolUnnormalized}[1]{D_{#1}}  %% unnormalized Sobol

\newcommand{\sobolClosed}[1]{\SobolNormalized{#1}^{\text{c}}}  %% closed normalized sobol
\newcommand{\sobolClosedD}[1]{\SobolUnnormalized{#1}^{\text{c}}}  %% closed unnormalized sobol
\newcommand{\totalVariance}{D}

\newcommand{\GSI}{\mathrm{GSI}^{\text{c}}}

\newcommand{\indOutputs}{\ell}
\newcommand{\inputVectorRandom}{X}  %% random input vector
\newcommand{\inputPointRandom}[1]{\inputVectorRandom_{#1}}
\newcommand{\outputVectorRandom}{Y}  %% random output vector
\newcommand{\probDist}{\mu}    %% probability distribution
\newcommand{\deterministicModel}{f}  %% deterministic model/code
\newcommand{\inputPoint}{x}  %% input point to a model
\newcommand{\DoE}{\mathcal{X}}   %% Design of experiments
\newcommand{\outputDeterministic}{y}  %% scalar deterministic output
\newcommand{\outputHighDimensional}{\outputDeterministic_{\indOutputs}} %% high-dimensional scalar deterministic output

\newcommand{\nPF}{N}  %% size of pick-freeze sample

\newcommand{\pfEstim}[1]{{\widehat{#1}}^{\mathrm{pf}}}  %% pick-freeze estimation hat

\newcommand{\indGPR}{j}
\newcommand{\GaussianProcess}{Z} 
\newcommand{\MeanFunction}{m}
\newcommand{\KernelFunction}{K}
\newcommand{\condGP}{\GaussianProcess_c}
\newcommand{\ConditionalMean}{\MeanFunction_c}
\newcommand{\ConditionalKernel}{\KernelFunction_c}
\newcommand{\condGPVector}{\GaussianProcess_n}
\newcommand{\ObservationVector}{\outputVectorDeterministic}
\newcommand{\KernelMatrix}{\KernelFunction(\DoE, \DoE)}
\newcommand{\KernelVector}[1]{\KernelFunction(\DoE, #1)}
\newcommand{\KernelVectorRow}[1]{\KernelFunction(#1, \DoE)}
\newcommand{\NumberGPite}{N_{Z}}
\newcommand{\nX}{N_X}

\newcommand{\indBasis}{q}
\newcommand{\coef}{c}   %% basis coefficients
\newcommand{\coefPred}{\hat{\coef}}
\newcommand{\basis}{v}  %% basis components
\newcommand{\basisVec}{\basis_{., \indOutputs}}   %% vector of basis components

\newcommand{\nbasis}{n_{b}}  %% number of components
\newcommand{\GramMatrix}{\mathbf{G}}

\newcommand{\depth}{h}
\newcommand{\depthThresh}{H_f}
\newcommand{\timeSteps}{t_n}
\newcommand{\numberTimeSteps}{N_t}
\newcommand{\velocityVec}{\vec{V}}
\newcommand{\velocityX}{u}
\newcommand{\velocityY}{v}
\newcommand{\gradientVec}{\vec{\nabla}}
\newcommand{\gravity}{g}
\newcommand{\gravityVec}{\vec{\gravity}}
\newcommand{\shearStress}{\tau_b}
\newcommand{\densityMixture}{\rho_m}
\newcommand{\hydraulicRad}{R}
\newcommand{\manning}{n}
\newcommand{\yieldStress}{\tau_c}
\newcommand{\plasticViscosity}{\mu_B}
\newcommand{\shearRate}{\dot{\gamma}}
\newcommand{\bottomWidth}{w}
\newcommand{\bottomElevation}{H_b}
\newcommand{\rightSlope}{z_1}
\newcommand{\leftSlope}{z_2}
\newcommand{\breachCoeff}{k}
\newcommand{\breachTime}{t_f}
\newcommand{\densityWater}{\rho_w}
\newcommand{\densitySolid}{\rho_s}
\newcommand{\volumetricConc}{C_v}
\newcommand{\floodProb}[1]{P_{f_{#1}}}
\newcommand{\arrivalTime}{T}

\newcommand{\maxFlowDepth}{H}

\begin{document}
\let\WriteBookmarks\relax
\def\floatpagepagefraction{1}
\def\textpagefraction{.001}

\shorttitle{Metamodel-Based Uncertainty and Sensitivity Analysis of Tailings Dam-Breach Flows}    

\shortauthors{Y.T. Sáo}  

\title [mode = title]{Metamodel-based methodology for uncertainty propagation and global sensitivity analysis of tailings dam-breach flows}


\author[1,2]{Yuri Taglieri Sáo}[orcid=0000-0003-2037-5451]
\cormark[1]

\ead{yuri.sao@unesp.br}

\credit{Conceptualization, methodology, investigation, software, validation, formal analysis, visualization, writing (original draft)}

\affiliation[1]{organization={Department of Mechanical Engineering, São Paulo State University ``Júlio de Mesquita Filho'' (UNESP)},%Department and Organization
            addressline={Av. Brasil, 56}, 
            city={Ilha Solteira},
            postcode={15385-000}, 
            state={São Paulo},
            country={Brazil}}

\affiliation[2]{organization={Department of Mathematics and Applications, Institut National des Sciences Appliquées (INSA)},
            addressline={135, Avenue de Rangueil}, 
            city={Toulouse},
            postcode={31077}, 
            state={Occitanie},
            country={France}}

\author[1]{Geraldo de Freitas Maciel}%[]

\credit{Conceptualization, writing (review and editing), supervision, funding acquisition}

\author[3]{Julian Cardoso Eleutério}

\credit{Data curation, investigation, writing (review and editing)}

\affiliation[3]{organization={Department of Hydraulic and Water Resources Engineering, Federal University of Minas Gerais (UFMG)},
            addressline={Escola de Engenharia}, 
            city={Belo Horizonte},
            postcode={31270-901}, 
            state={Minas Gerais},
            country={Brazil}}

\cortext[1]{Yuri Taglieri Sáo}


\begin{abstract}
Tailings dam-breach analyses are essential for flood-hazard assessment, emergency planning and risk estimation, but their results are strongly affected by uncertainties in breach development, released volume and tailings rheology. This study proposes an efficient probabilistic methodology that integrates uncertainty quantification and global sensitivity analysis for tailings dam-breach studies. High-dimensional outputs (spatial maps) and the computational cost of deterministic simulations are addressed through dimensionality reduction and metamodeling. The methodology is demonstrated on a benchmark case with complex terrain using HEC-RAS v6.6 and considering uncertainties in breach parameters and rheological properties. The results quantify uncertainty in maximum flow depth and arrival time, characterize their statistical distributions and identify the spatial influence of the main input variables through sensitivity maps. Sensitivity indices reveal the dominance of breach parameters near the dam and yield stress farther downstream. The modular and non-intrusive framework can be coupled with other deterministic models and applied to different dam-breach scenarios, supporting more standardized and risk-informed assessments.
\end{abstract}

\begin{graphicalabstract}
\includegraphics{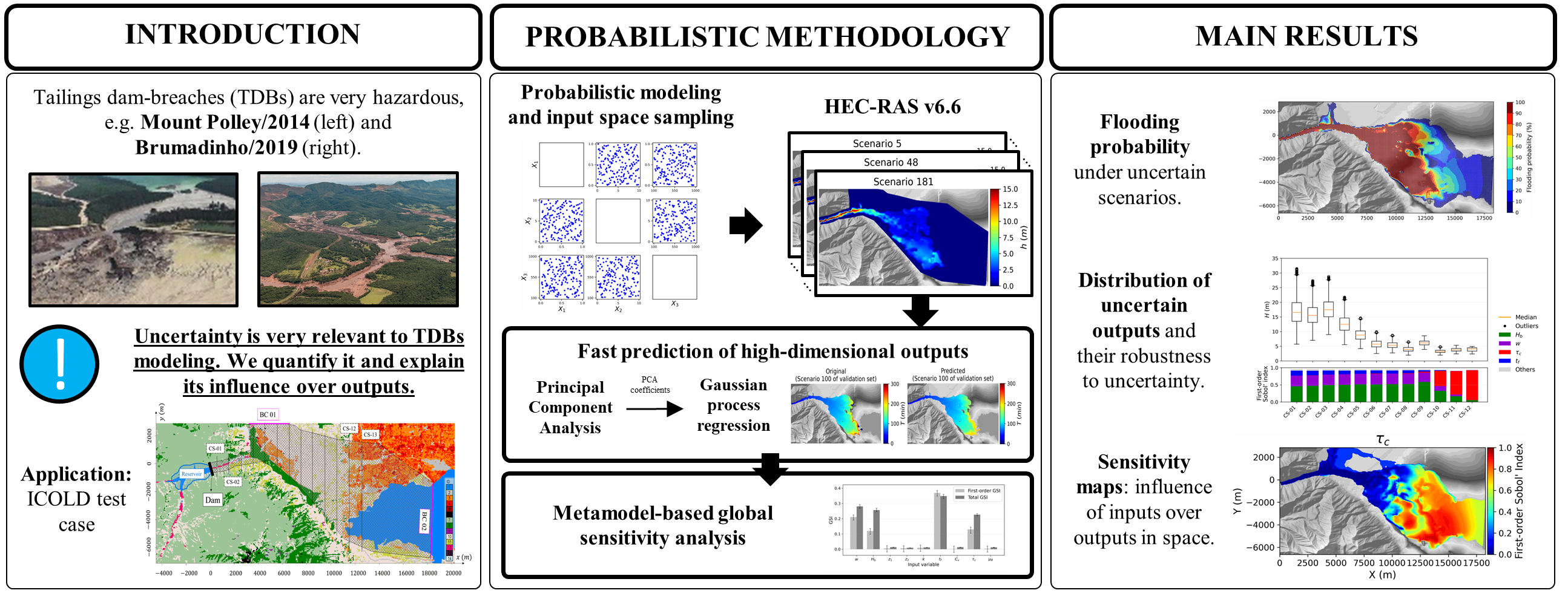}
\end{graphicalabstract}

\begin{highlights}
\item Probabilistic modelling of tailings dam-breach flow is proposed.
\item PCA and Gaussian processes enable fast prediction of flow maps.
\item Uncertainty is quantified for depth and arrival time outputs.
\item Sensitivity analysis identifies influential inputs over space.
\item The framework supports hazard assessment under epistemic uncertainty.
\end{highlights}

% Keywords

\maketitle

% Main text
\section{Introduction}

Tailings storage facilities (TSFs) are large structures used to store the waste generated during mineral processing \citep{Blight2009}. Like any engineered structure, TSFs are susceptible to failure and may pose substantial risks to public safety, the environment and infrastructure \citep{Rico2008a, Kossoff2014}, depending on their size, their content and the vulnerability of downstream areas. In fact, an average of 2.5 failures per year has been reported since 1915 \citep{Lyu2019, Piciullo2022}. Recent disasters highlight the high level of risk associated with these TSFs, such as the Mount Polley dam breach (Canada/2014) \citep{Petticrew2015}, the Fundão dam breach (Brazil/2015), or ``Mariana's disaster'' \citep{Carmo2017}, and the Córrego do Feijão dam rupture (Brazil/2019), or ``Brumadinho's disaster'' \citep{Rotta2020, Lumbroso2021}. We refer to \citep{Rico2008b, Kossoff2014, Rana2022} for a more comprehensive overview of tailings dam-breach incidents. The number of incidents involving tailings dams over the years and this series of widely publicised disasters have pressured mining industry stakeholders to reduce both the frequency and consequences of such failures. 

In this context, the consequences of a possible breach must be evaluated through studies known as tailings dam-breach analyses (TDBA) \citep{Martin2019, Gildeh2021}. These analyses are similar to standard water-retaining dam-breach analyses \citep{Gildeh2021}, in which a deterministic model solves the flood-routing problem and the resulting outputs are used to estimate relevant flow properties, such as flooded area, maximum flow depth and maximum velocity. These outputs are useful for tailings dam design, emergency planning and safety management \citep{Martin2019, Paiva2020}. Risk can then be estimated by also considering exposure and vulnerability \citep{Ward2020}. Although general guidelines on the procedures to perform TDBAs are available in the literature, specific standardization for the geotechnical/hydrodynamical propagation of tailings is yet to be developed \citep{Martin2019, Liu2020_dam}. Therefore, the lack of standardization can potentially introduce subjectivity in the TDBAs, which may impact risk estimation and introduce uncertainty into risk-reduction strategies and decision-making processes.

Several sources of uncertainty are associated with dam-breach analyses in general, such as the trigger mechanism of the failure and the development of the breach \citep{Tian2021, Walsh2021, DaSilva2023}. For TDBAs specifically, the released volume \citep{Rico2008a, Ghahramani2020} and rheological behavior of tailings \citep{Boger2013, Moon2019, Piette2022} are particularly important. We refer to \citep{Gildeh2021, Ghahramani2024} for a more detailed description of uncertainties related to TDBAs. In particular, much of the uncertainty is epistemic because of limited knowledge of the underlying physical processes and the difficulty of determining uncertain properties precisely \citep{Beven2018}. In a forward-simulation context, input uncertainty may propagate nonlinearly to the outputs of interest to the analysis, due to the complexity of the deterministic models. Consequently, two common practices in TDBAs, i.e. worst-case analysis and one-at-a-time variation, may be unrealistic and inconsistent or provide only local results that depend on the selected nominal values \citep{Tarantola2002, Wagener2019}. This may lead to overly conservative and/or unrealistic risk estimation.

To account for uncertainty in these studies, uncertainty quantification approaches provide useful tools towards the development of a methodology for TDBAs that treats input uncertainty statistically. Among many techniques, non-intrusive methods, such as the Monte Carlo method, are useful since they do not require modifications in the deterministic code and are easily parallelizable \citep{Cunha2014}. These techniques allow input variables to be modelled probabilistically by assigning probability density functions (PDFs). A set of possible scenarios, or combinations of input variables, is generated systematically using statistical sampling techniques, e.g. pure Monte Carlo sampling, Latin hypercube sampling (LHS), Sobol' sequences, etc \citep{Pronzato2012}. By running all scenarios in a deterministic code, input uncertainty is propagated to outputs, generating output distributions and enabling appropriate statistical analysis of the outputs \citep{Soize2017, Ghanem2017}. In addition, the influence of uncertain input variables on the outputs can be measured by employing global sensitivity analysis (GSA) \citep{Saltelli2008, DaVeiga2021}, typically used to identify important input variables, rank their importance, fix values to unimportant variables and detect possible interactions. These capabilities support the explanation of both the phenomenon and the effects of uncertainty in terms of statistical indices.

Applications of probabilistic approaches are widely explored in the literature to study environmental problems \citep{Zhao2020, Zegers2020, Alipour2022, Rohmer2022, Elgarroussi2022, Huang2023, Houdard2023}. The use of a probabilistic approach to investigate tailings dam-breach flows poses two methodological challenges: (i) the simulation code is relatively expensive; and (ii) the outputs are high-dimensional (spatial maps or time series). To deal with the first challenge, the usage of a metamodeling technique is crucial, where few expensive data points are used to train a fast-to-evaluate statistical model, e.g. Gaussian process regression \citep{Williams1995}, polynomial chaos expansion \citep{Sudret2008}, artificial neural networks \citep{Fonseca2003}, among others. To address the second challenge, a dimensionality reduction technique is employed, approximating the outputs through a truncated basis expansion, forming a reduced space that represents outputs with significantly fewer variables, e.g. principal component analysis (PCA) \citep{Abdi2010}, weighted-PCA \citep{Chaleshtori2024}, wavelets \citep{Perrin2021}, among others. As demonstrated in \citep{Li2020, Sao2025}, by combining dimensionality reduction and metamodeling, both the uncertainty propagation to high-dimensional outputs and the estimation of sensitivity indices over the entire domain, i.e. sensitivity maps (SMs) \citep{Marrel2011, Li2020, Sao2025}, are now non-prohibitive. As shown in this work, SMs are useful to evaluate \textit{where} an output is sensitive to a given input variable; this information is unavailable while evaluating outputs in a given point of space or aggregated measures (e.g. flooded area). 

In the TDBA context, however, general probabilistic methods remain largely unexplored, except for a few studies \citep{Melo2023, Ghahramani2024}. To the authors' knowledge, uncertainty propagation and GSA have not yet been jointly applied to high-dimensional outputs from TDBA models. Although the individual methods and their integration into probabilistic frameworks have been extensively investigated in other environmental modelling contexts \citep{Rohmer2016, Elgarroussi2022, Sheikholeslami2025}, their adaptation and combined application to TDBAs remain unexplored. The main contribution of this study is therefore the adaptation and application of an integrated and non-intrusive framework for the probabilistic analysis of tailings dam-breach flows. The framework combines established techniques for dimensionality-reduction (PCA) and metamodeling (GPR) to enable efficient predictions of quantities of interest, together with variance-based global sensitivity analysis to quantify both aggregated and spatially distributed input effects. A relevant benchmark case \citep{Zenz2013} is used to illustrate results of the probabilistic analysis, characterized by complex topography and by hydraulic and land-cover features representative of real-world dam-breach settings. Its application provides three complementary types of information that are generally unavailable from conventional deterministic analyses: flooding-probability maps, statistical distributions of quantities of interest and SMs identifying how the influence of uncertain parameters varies across the flooded domain. The framework is not restricted to the outputs considered in this study and may also be applied to other hazard-related quantities, e.g. maximum flow velocity and shear stress. These capabilities address the growing demand for uncertainty and sensitivity analyses in TDBAs \citep{Martin2019, Gildeh2021, Ghahramani2024}, while overcoming the limitations of worst-case and one-at-a-time analyses. Therefore, this work helps fill the gap in the probabilistic analysis of tailings dam-breach flows and demonstrates an integrated framework for conducting TDBAs under uncertainty, supporting more systematic, transparent and risk-informed engineering assessments.

This work is organized as follows. First, in section \ref{sec:deterministic_model}, we justify and describe the choice of the deterministic model and present a test case. The resulting data is used to feed the probabilistic framework presented in section \ref{sec:probabilistic_framework}, where uncertainty propagation and GSA concepts are detailed. Then, results are presented and discussed in section \ref{sec:results}, followed by the conclusions in section \ref{sec:conclusion}.

\section{Deterministic approach}
\label{sec:deterministic_model}

The deterministic approach is essential to provide accurate data for the probabilistic framework. The numerical deterministic model must be able to simulate the main physical features of the tailings flows and account for the particularities of the phenomenon, e.g. rheological and physical characteristics of tailings and breaching mechanisms of tailings dams. We refer to \citep{Martin2019, Gildeh2021, Sreekumar2024} for more details about the particularities of tailings dam-breach flows. According to the CDA classification presented by \citep{Martin2019}, the scope of this work is the evaluation of tailings dam-breach flows with low to medium volumetric concentrations. Next, we present the chosen numerical model and the respective assumptions.

% The analysis of hypothetical tailings dam-breach flows is an important part of TDBAs. Briefly, some sequential steps are considered: hydrological studies to define the watershed; definition of failure mode; analysis of the breach and mobilized volume; and propagation of tailings over a complex terrain \citep{Paiva2020, Gildeh2021, Sreekumar2024}. Analysts must ensure that the main physical features of tailings flows are properly modeled into the deterministic simulation model. Therefore, the choice of a deterministic model is crucial.

\subsection{Deterministic model for simulating tailings dam-breach flows}

Deterministic simulations of tailings dam-breach flows are similar to standard flood routing, where an inlet boundary condition is prescribed to reflect the release of volume, followed by the propagation of the flow throughout the domain \citep{Sreekumar2024}. In the literature on tailings flows, two main hypotheses are assumed: the one-phase approach and the shallow-water hypothesis. First, concerning the one-phase approach, numerical solvers usually treat the flowing material as a single continuum fluid, where rheological behavior and solid-liquid interactions are described by a single rheological law \citep{Paiva2020, Sreekumar2024}. One-phase approaches generally provide satisfactory results for global macroscopic properties of the flow, such as flow depth and velocity field, thus representing the main physical features of the flow \citep{Zegers2020, Gibson2021, Lumbroso2021}. This assumption is common in other mass-movement applications with low to medium volumetric concentration, such as mudfloods, mudflows and debris flows to some extent \citep{Obrien1988, Pirulli2017, Gibson2022}. For more concentrated mass movements, multi-phase models may be more appropriate, capturing additional processes, e.g. solid-fluid relative motion, segregation and pore pressure development; however, these cases are out of the scope of this work. Second, the shallow-water hypothesis is employed when the characteristic depth is significantly smaller than the other length scales, allowing substantial simplification of the model while still reproducing macroscopic flow properties satisfactorily. It is widely used in flood routing applications and in the tailings dam-breach literature \citep{Ligier2020, Lumbroso2021, Gibson2021, Melo2023}.

Based on these assumptions, HEC-RAS v6.6 was selected to perform the deterministic simulations of tailings dam-breach flows. It solves the shallow-water equations (SWE) using the finite-volume method to model the fluid dynamics which must satisfy the hypotheses of incompressible fluid and shallow-water condition. More details can be found in the HEC-RAS documentation \citep{Hecras66}. By neglecting effects of porosity, wind, wave forcing and Coriolis effect, SWE system is given by:
\begin{equation}
\label{eq:continuity}
\begin{gathered}
\frac{\partial \eta}{\partial t} + \gradientVec \cdot (\depth \velocityVec) = 0 \\
\frac{\partial \velocityVec}{\partial t} + (\velocityVec \cdot \gradientVec)\velocityVec
= -\gravityVec \, \cos^2 \phi \, \gradientVec \eta
+ \frac{1}{\depth}\gradientVec \cdot (\nu_t \depth \gradientVec\velocityVec)
- \frac{\shearStress}{\densityMixture \hydraulicRad} \frac{\cos \psi}{\cos \phi}\frac{\velocityVec}{|\velocityVec|}
\end{gathered}
\end{equation}
where $\eta$ is the flow surface elevation, $\depth$ is the free-surface elevation in relation to the terrain (or depth), $\phi$ is the water surface slope, $\psi$ is the inclination angle of the velocity direction, $\velocityVec = (\velocityX,\velocityY)$ is the velocity vector, $\gravityVec$ is the vector of gravity acceleration, $\nu_t$ is the turbulent viscosity, $\shearStress$ is the total basal shear stress, $\densityMixture$ is the mixture density and $\hydraulicRad$ is the hydraulic radius. The mixture density $\densityMixture$ is a function of the volumetric concentration $\volumetricConc$, as follows:
\begin{equation}
\label{eq:density_mixture}
    \densityMixture = \densityWater + (\densitySolid - \densityWater) \volumetricConc
\end{equation}
where $\densityWater = 998 \ \mathrm{kg.m^{-3}}$ is the water density and $\densitySolid = 2650 \ \mathrm{kg.m^{-3}}$ is the solid-particle density.

The basal shear stress $\shearStress$ can be divided into two components. The first component is related to the roughness of the terrain, which is represented by the Manning coefficient $\manning$. The second component is the shear stress related to the internal friction of the fluid and directly linked to the fluid rheology; here, we use the Bingham rheological model, under the hypothesis of simple shear. The basal shear stress is written as:
\begin{equation}
\label{eq:basal_shear_stress}
    \shearStress = \frac{\densityMixture \gravity \manning^2}{\hydraulicRad^{1/3}} |\velocityVec|^2 + \yieldStress + \plasticViscosity \shearRate
\end{equation}

where $\yieldStress$ is the yield stress, $\plasticViscosity$ is the plastic viscosity and $\shearRate$ is the shear rate. Among many closure expressions for the shear rate \citep{Pastor2015, Sao2023}, HEC-RAS v6.6 employs the following expressions as a function of the mean velocity $\bar{V}$ and $\depth$:
\begin{equation}
\label{eq:shear_rate}
\shearRate = \frac{3 \bar{V}}{\depth}
\end{equation}

% In HEC-RAS v6.6, $\shearRate$ is given by Eq. \ref{eq:shear_rate}, which is an approximated closure equation for the model by considering a zeroth-order estimation of shear rate, i.e. derived from fully developed conditions of a non-Newtonian fluid flow. These assumptions are not strictly satisfied, since tailings dam-breach flows are transient and non-uniform. Therefore, Eq. \ref{eq:shear_rate} must be interpreted as a simplifying modelling choice. Furthermore, the exact calculation of a zeroth-order approximation of shear rate for non-Newtonian rheological models requires the velocity profile of the flow in fully developed conditions, turning the shear rate a function of rheological parameters and terrain slope, as shown in \citep{Sao2023, Muchiri2025}.

\begin{figure}[H]
    \centering
    \includegraphics[width=0.6\linewidth]{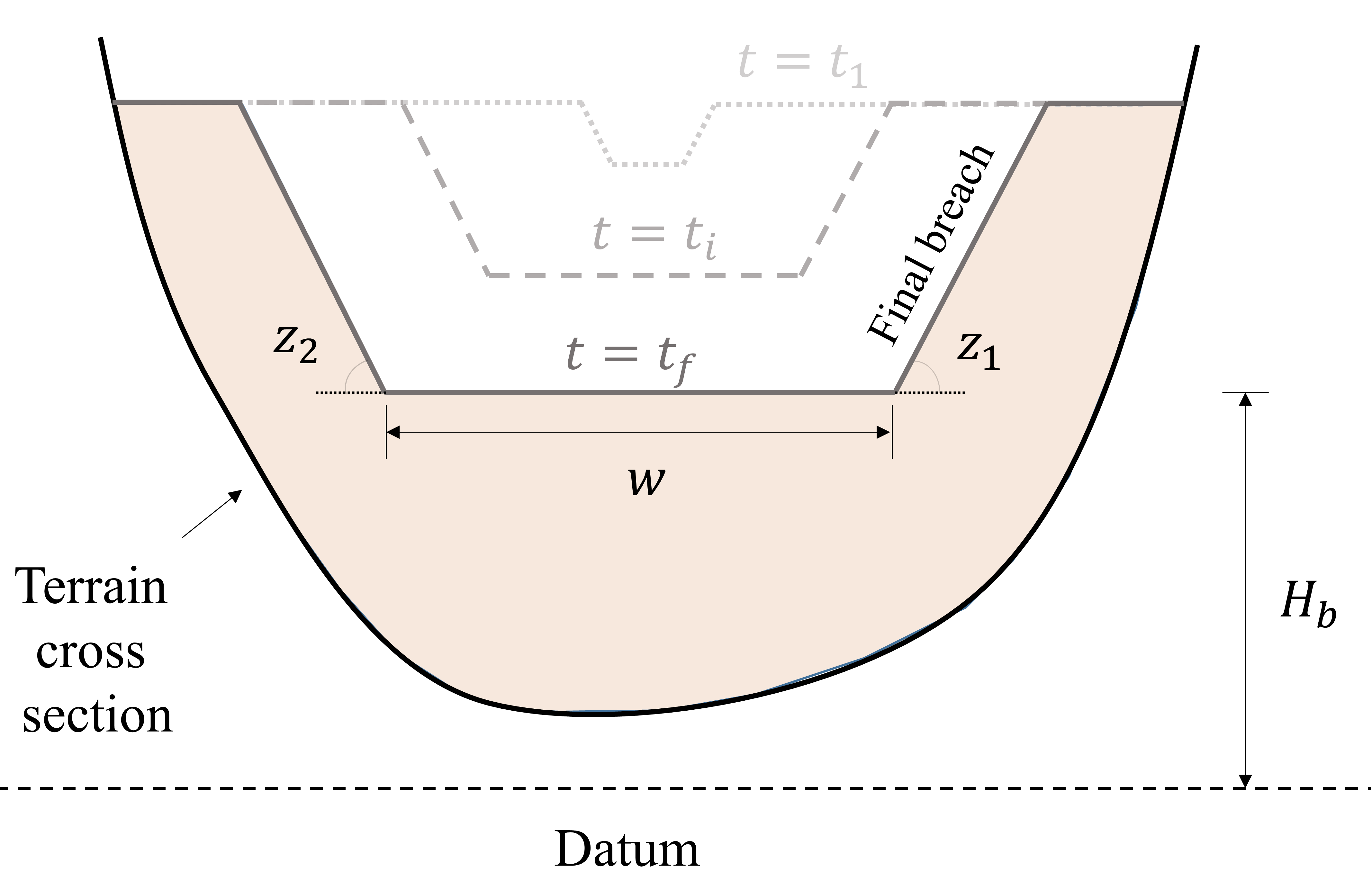}
    \caption{Schematic of the dam cross section and breached area. Parameters related to the breached area are shown.}
    \label{fig:schematic_breach}
\end{figure}

At the inlet boundary, breach development is represented by a trapezoidal section generated by overtopping of the dam. The breached area develops over time following a sinusoidal law, with a growth ratio in the vertical and horizontal directions of $1$, starting at the dam crest and parameterized by the following variables: final bottom width $\bottomWidth$, final bottom elevation $\bottomElevation$, left slope $\leftSlope$, right slope $\rightSlope$, breach weir coefficient $\breachCoeff$ and breach formation time $\breachTime$ (see Fig. \ref{fig:schematic_breach}). While $\breachTime$ controls the timing of the released hydrograph, other parameters ($\bottomWidth, \bottomElevation,\rightSlope, \leftSlope$) determine the trapezoidal breached area. Additionally, $\bottomElevation$ controls the volume released from the upstream reservoir, since upstream material below the breach bottom cannot be released downstream. We note that breach development is a complex phenomenon, which depends on the breach location, material properties, degree of saturation and current condition of the dam \citep{Tian2021, Walsh2021}. Therefore, the mechanism adopted herein is a representative model of a possible breach; if a more accurate model were available, it could be incorporated into the methodology by changing the breach-related input parameterization.

\begin{figure}[H]
    \centering
    \includegraphics[width=0.95\linewidth]{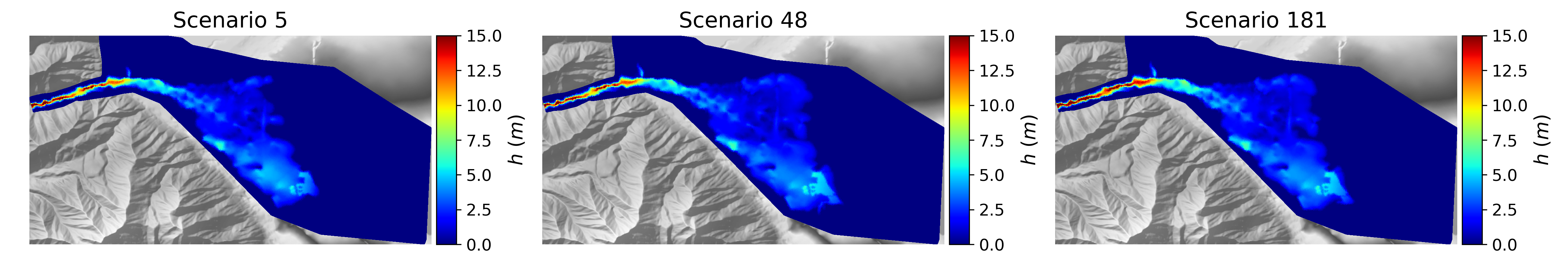}
    \caption{Examples of maps of maximum flow depth in different simulation scenarios.}
    \label{fig:examples_maximum_flow_depth}
\end{figure}

\begin{figure}[H]
    \centering
    \includegraphics[width=0.95\linewidth]{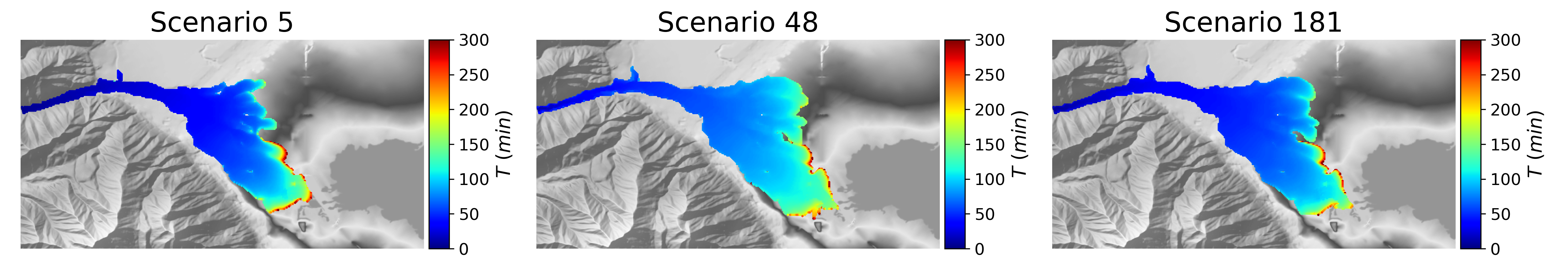}
    \caption{Three examples of arrival times maps in different simulation scenarios.}
    \label{fig:examples_arrival_times}
\end{figure}

By solving the SWE system, velocity and depth over space and time are obtained. Let $\depth_{\indOutputs}(\timeSteps)$ be the flow depth field, where $\{ \timeSteps \}_{n=0}^{\numberTimeSteps}$ is the discretized time in $\numberTimeSteps$ time steps and $\indOutputs = 1, \dots, \nOutputDimensions$ is the index of $\nOutputDimensions$ cells in the discretized spatial domain. The main outputs of interest are the maximum flow depth $\maxFlowDepth$ and arrival times $\arrivalTime$, using the same indexing convention as $\depth_{\indOutputs}$. The maximum flow depth map is defined as the maximum water depth obtained over all discrete solver time steps; the arrival time is the first instant at which the flow depth exceeds the flooding threshold $\depthThresh$. Both outputs are calculated as given by Eq. \ref{eq:max_flow_depth}. Here, we consider $\depthThresh = 0.31 \ m$. Figures \ref{fig:examples_maximum_flow_depth} and \ref{fig:examples_arrival_times} show examples of maps of maximum flow depth and arrival times, respectively.
\begin{equation}
\label{eq:max_flow_depth}
    \maxFlowDepth_{\indOutputs} = \max_{0 \leq n \leq \numberTimeSteps} \depth_{\indOutputs}(\timeSteps) \quad ; \quad \arrivalTime_{\indOutputs} =
\begin{cases}
    \min\{ \timeSteps \}, & \text{if} \ \depth_{\indOutputs}(\timeSteps) > \depthThresh   \\
    \text{NaN}, & \text{otherwise}
\end{cases}
\end{equation}

\subsection{Test case and numerical settings}
\label{subsec:test_case}

In this work, we apply the probabilistic methodology to a test case developed by the ICOLD (International Commission on Large Dams), referred to as ``ICOLD test case'' herein \citep{Zenz2013}. It consists of a hypothetical dam surrounded by a fictional complex terrain, with specified land uses and land covers. The ICOLD test case has been used in several studies, including water dam-breach analyses \citep{Ahmadisharaf2016, Peter2018, DaSilva2023} and tailings dam-breach analyses \citep{Melo2023}. Since it is a hypothetical case, we can explore different conditions and sources of uncertainty, such as breach parameters and rheological parameters. 

\begin{figure}[H]
    \centering
    \includegraphics[width=0.95\linewidth]{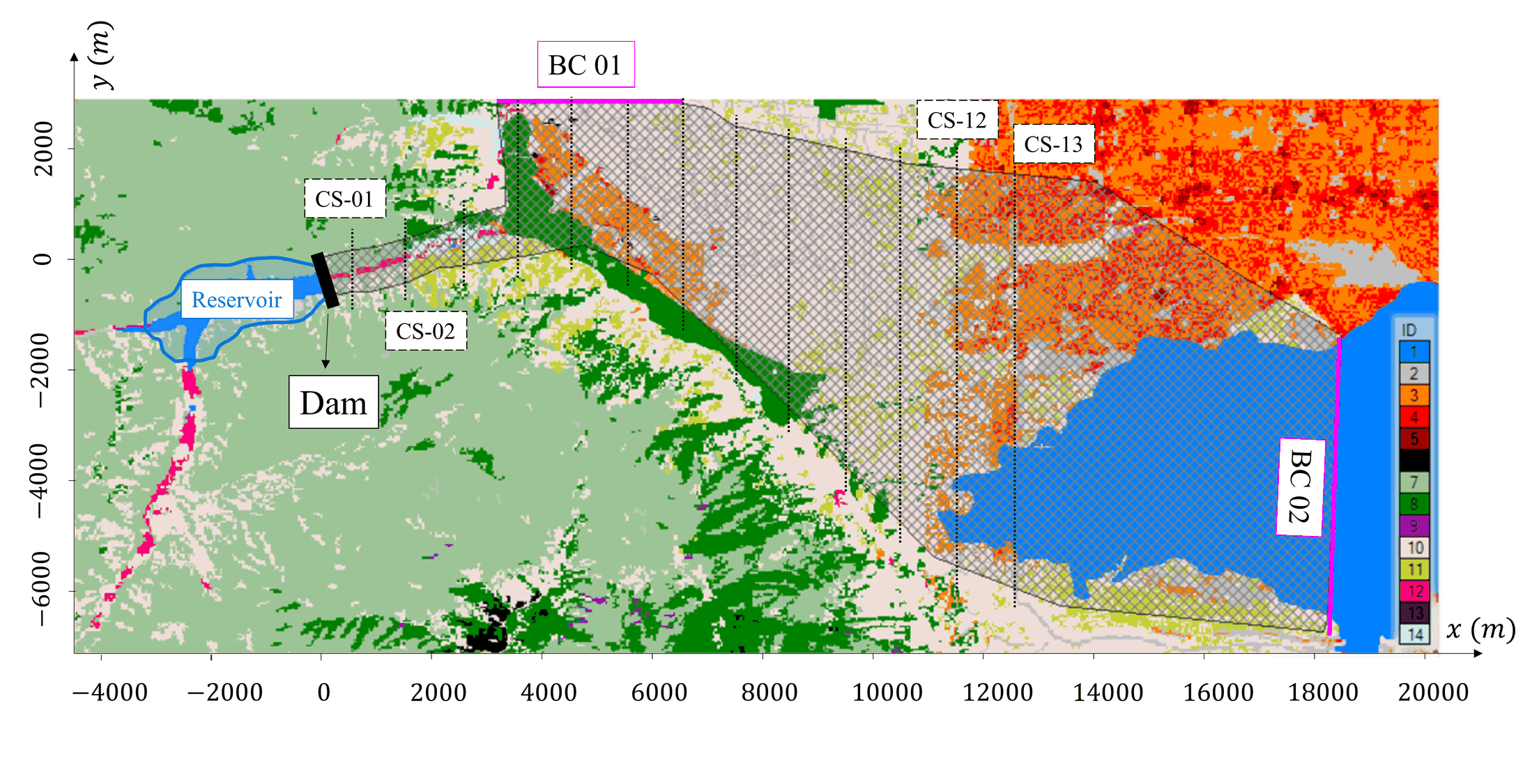}
    \caption{Schematic of the ICOLD test case: location of the dam, land use, land cover and computational domain. Fourteen typologies represented by the IDs in the figure were considered. Typology, ID and Manning coefficient follows. ID-1: open water (0.019); ID-2: open space developed areas (0.035); ID-3: low-intensity open space developed areas (0.065); ID-4: medium-intensity open space developed areas (0.074); ID-5: high-intensity open space developed areas (0.070); ID-6: barren land (0.056); ID-7: deciduous forest (0.138); ID-8: evergreen forest (0.158); ID-9: mixed forest (0.123); ID-10: shrubs (0.057); ID-11: grasslands (0.038); ID-12: pasture (0.070); ID-13: cultivated cropland (0.068); ID-14: woody wetlands (0.155)}
    \label{fig:hypothetical_dam_terrain}
\end{figure}

Figure \ref{fig:hypothetical_dam_terrain} shows a schematic containing information about the ICOLD test case. The digital elevation model (DEM) has a horizontal resolution of $9.5 \ \mathrm{m}$. The hypothetical dam is $61 \ \mathrm{m}$ high, with crest elevation and length of $272$ and $360$ meters, respectively. The upstream reservoir formed by the dam stores $38 \times 10^6 \ \mathrm{m^3}$ of material. The downstream terrain forms a confined V-shaped valley along the first $3.6 \ \mathrm{km}$, before opening onto a flat region with different land uses and land covers, such as urban areas, forests and cultivated lands. Figure \ref{fig:hypothetical_dam_terrain} details the land uses and land covers, presenting the corresponding Manning coefficients of each land classification. 

The following assumptions were adopted in the tailings dam-breach analysis of this work: a gradual overtopping-induced breach and mining tailings as the stored and released material. The first hypothesis considers a trapezoidal breached section that starts at the crest and develops gradually, following a sinusoidal development over time. Further information is provided by \citep{Hecras66, Xu2009, Froehlich2016}. The second hypothesis states that the released material is composed of mining tailings, modeled as non-Newtonian viscoplastic fluids, following the Bingham rheological model \citep{Boger2013}. Therefore, the material composition, including its origin, grain-size distribution, and mineralogy, is not specified explicitly. Instead, the released material is characterized through its density and rheological properties.

The computational mesh covers an area of approximately $70 \ \mathrm{km^2}$, containing $27680$ cells with dimensions of $55 \times 55 \ \mathrm{m}$. No significant differences were observed when finer meshes were used. Two downstream boundary conditions are prescribed as normal-depth boundary conditions, with slopes of $0.005 \ \mathrm{m/m}$ for the north boundary of the terrain and $0.0001 \ \mathrm{m/m}$ for the downstream lake, as indicated by Fig. \ref{fig:hypothetical_dam_terrain}. The total simulation time was $5$ hours ($300$ minutes) and the time discretization followed the Courant-Friedrichs-Lewy (CFL) condition, with maximum CFL number of $0.8$. A maximum of 40 iterations per time step was used. No significant differences were observed when finer time steps or a larger maximum number of iterations were used.

\section{Probabilistic framework}
\label{sec:probabilistic_framework}

The probabilistic framework applied in this work is detailed in this section. Here, a Monte Carlo probabilistic method is used to propagate uncertainty from inputs to outputs, since it is non-intrusive and easily parallelizable \citep{Cunha2014}. First, probabilistic modeling of input variables is described in subsection \ref{subsec:probabilistic_modeling}. Next, subsection \ref{subsec:UQ} presents dimensionality reduction and metamodeling techniques to predict accurate data efficiently. Finally, in subsection \ref{subsec:GSA}, variance-based global sensitivity analysis (GSA) techniques to estimate sensitivity indices are described. We refer to \citep{Rohmer2016, Nagel2020, Li2020, Sao2025} for similar applications of probabilistic frameworks. Although the methodology is demonstrated here for maximum flow depth and arrival time, it can be applied to other spatially distributed hazard-related quantities, e.g. maximum flow depth and shear stress at the bottom, provided that the corresponding output maps are generated by the deterministic model.

\subsection{Probabilistic modeling of input variables for the ICOLD test case}
\label{subsec:probabilistic_modeling}

The deterministic model includes two main sources of uncertainty. The first is related to the breach parameters: $\bottomElevation$, $\bottomWidth$, $\rightSlope$, $\leftSlope$, $\breachTime$ and $\breachCoeff$. The second is related to the physical properties of the material, which is modeled as a Binghamian fluid: $\yieldStress$, $\plasticViscosity$ and $\volumetricConc$. Uncertain input variables are treated as random input variables, described by the random input vector $\inputVectorRandom=\{\bottomElevation, \bottomWidth, \rightSlope, \leftSlope, \breachTime, \breachCoeff, \yieldStress, \plasticViscosity, \volumetricConc \}$. Input variables are considered to be mutually independent, as shown by \citep{DaSilva2023} for breach parameters. 

To account for input uncertainty, probability density functions (PDFs) are assigned to the input variables, denoted as $\probDist_X$, and must satisfy physical constraints (e.g. positivity) and maximum entropy condition \citep{Soize2017}. By following the maximum entropy principle, we avoid adding unjustified information to the probabilistic modeling. Although probability distributions for breach parameters have been proposed in the literature \citep{DaSilva2023}, specific data on breach development and tailings rheology for tailings dam-breach events are scarce. Thus, uniform distributions were adopted to represent epistemic uncertainty, ensuring the most unbiased distribution to the probabilistic modeling. The main assumption of a uniform distribution is that every scenario has the same probability of occurring, which is a reasonable consideration when prior information is scarce. Table \ref{tab:PDFs} lists the random input variables, their respective PDFs, and the rationale for the selected statistical supports. 

\begin{table}[h!]
\centering
\caption{Probabilistic modeling of random input variables and justification of statistical supports.}
\label{tab:PDFs}
\small
\begin{tabularx}{\textwidth}{l l l X}
\toprule
\textbf{Input variable} & \textbf{PDF and support} & \textbf{Unit} & \textbf{Justification of support} \\
\midrule
$\bottomElevation$  & $\mathcal{U}\!\left[220,\, 250 \right]$ & m 
& Geometry constraint. \\

$\bottomWidth$      & $\mathcal{U}\!\left[0.01,\, 100\right]$ & m 
& Geometry constraint. \\

$\rightSlope$       & $\mathcal{U}\!\left[0.1,\, 2.0\right]$  & m/m 
& Geometry constraint. \\

$\leftSlope$        & $\mathcal{U}\!\left[0.1,\, 2.0\right]$  & m/m 
& Geometry constraint. \\

$\breachTime$       & $\mathcal{U}\!\left[0.08,\, 2.00\right]$ & h 
& Data from \citep{Froehlich2008, Froehlich2016} and total time of simulation constraint. \\

$\breachCoeff$      & $\mathcal{U}\!\left[1.1,\, 1.8\right]$  & -- 
& Expert judgement from \citep{Hecras66} \\

$\yieldStress$      & $\mathcal{U}\!\left[1.0,\, 200.0\right]$ & Pa 
& Data (see Fig. \ref{fig:support_yield}) \\

$\plasticViscosity$ & $\mathcal{U}\!\left[0.1,\, 15.0\right]$  & Pa$\cdot$s 
& Data (see Fig. \ref{fig:support_viscosity}) \\

$\volumetricConc$   & $\mathcal{U}\!\left[20,\, 60\right]$    & \% 
& Data (see Figs. \ref{fig:support_yield} and \ref{fig:support_viscosity}) and expert judgement \citep{Martin2019, Liu2020_dam, Sreekumar2024} \\
\bottomrule
\end{tabularx}
\end{table}

Input scenarios are sampled from the joint input distribution, which are evaluated by the deterministic code. The sampling technique Latin-hypercube sampling (LHS) \citep{Pronzato2012} is employed to sample the input space. The input sample is then divided into two sets: (i) training set, used to train metamodels, and (ii) validation set, to validate the metamodels independently.

\subsection{Fast and accurate prediction of high-dimensional outputs}
\label{subsec:UQ}

To address the high dimensionality of the outputs and the high computational cost of each deterministic simulation, linear basis-expansion and metamodeling techniques are used, respectively. Other works have applied similar approaches \citep{Elgarroussi2022, Rohmer2023, Sheikholeslami2025}. The main idea is to expand output data into a reduced-order space, effectively reducing the dimensionality of the problem, and then to construct metamodels for the reduced-basis coefficients. The metamodeling step allows new, previously unexplored basis coefficients to be predicted quickly and accurately, which can be projected back to the original space by using the reduced-order basis vectors, thereby producing a new high-dimensional map. Once the metamodels have been trained, additional output maps can therefore be generated without further evaluations of the computationally expensive deterministic simulator. In this work, principal component analysis (PCA) and Gaussian process regression (GPR) are used for linear basis-expansion and metamodeling, respectively. The sequential framework of PCA and GPR was implemented in Python.

\subsubsection{Principal component analysis as linear basis-expansion technique}
\label{subsubsec:PCA}

\newcommand{\dataMatrixPCA}{\mathcal{Y}}
\newcommand{\centeredData}{\widetilde{\dataMatrixPCA}}
\newcommand{\covarianceMatrixPCA}{\Sigma}
\newcommand{\coefficientsMatrixPCA}{C}
\newcommand{\basisMatrixPCA}{V}
\newcommand{\dimensionMean}{\overline{\outputPoint}}
\newcommand{\outputPoint}{y}

Principal component analysis (PCA) is a data-driven dimensionality-reduction technique, which provides a linear representation of high-dimensional data in a reduced space. For more details, we refer to \cite{Abdi2010}. Let $\dataMatrixPCA = \left[ \outputPoint^{(1)}, \dots, \outputPoint^{(\nDoE)} \right]^{\top} \in \R^{\nDoE \times \nOutputDimensions}$, where $\outputPoint^{(\indPoints)} = \outputPoint(\inputPoint^{(\indPoints)})$ is the $\indPoints$-th high-dimensional output obtained by the deterministic simulator. The dimension-wise mean $\dimensionMean \in \R^{\nOutputDimensions}$ is calculated as:
\begin{equation}
\label{Eq:dimension_wise_mean}
    \dimensionMean = \frac{1}{\nDoE} \sum^{\nDoE}_{\indPoints=1} \outputPoint^{(\indPoints)}
\end{equation}

Centered data $\centeredData \in \R^{\nDoE \times \nOutputDimensions}$ is obtained by subtracting $\overline{\outputPoint}$ from each row of $\dataMatrixPCA$. Then, the empirical covariance matrix $\covarianceMatrixPCA \in \R^{\nOutputDimensions \times \nOutputDimensions}$ is calculated based on $\centeredData$:
\begin{equation}
\label{Eq:covariance_matrix_PCA}
    \covarianceMatrixPCA = \frac{1}{\nDoE - 1} \centeredData^{\top} \centeredData
\end{equation}

The eigenvalue problem $\covarianceMatrixPCA \basis_{\indBasis} = \lambda_{\indBasis} \basis_{\indBasis}$ is solved and the eigenvalues are sorted in decreasing order $\lambda_1 \geq ... \geq \lambda_{\nOutputDimensions}$, representing the data variance explained by the corresponding eigenvectors. The eigenvector associated with the $\indBasis$-th eigenvalue is denoted by $\basis_{\indBasis} = (\basis_{\indBasis,1}, \dots, \basis_{\indBasis, \nOutputDimensions})^{\top} \in \R^{\nOutputDimensions}$. The truncated PCA basis is obtained by retaining the first $\nbasis$ eigenvectors, with $\nbasis \ll \nOutputDimensions$, and is denoted by $\basisMatrixPCA = (\basis_{\indBasis, \indOutputs})_{1 \leq \indBasis \leq \nbasis, \ 1 \leq \indOutputs \leq \nOutputDimensions} \in \R^{\nbasis \times \nOutputDimensions}$, where the rows of $\basisMatrixPCA$ contain the retained basis vectors. The Gram matrix can be defined as $\GramMatrix = \basisMatrixPCA \basisMatrixPCA^{\top}$. The PCA basis coefficients are obtained by projecting the centered data onto the truncated basis $\coefficientsMatrixPCA = \centeredData \basisMatrixPCA^{\top} \in \R^{\nDoE \times \nbasis}$. Equivalently, $\coefficientsMatrixPCA = ( \coef_{\indBasis}(\inputPoint^{(\indPoints)}) )_{1 \leq \indPoints \leq \nDoE, \ 1 \leq \indBasis \leq \nbasis}$, where the $\indPoints$-th row of $\coefficientsMatrixPCA$ contains the basis coefficients associated with the
input scenario $\inputPoint^{(\indPoints)}$: $\coef(\inputPoint^{(\indPoints)}) = (\coef_1(\inputPoint^{(\indPoints)}),\ldots,\coef_{\nbasis}(\inputPoint^{(\indPoints)}))^\top
\in \R^{\nbasis}$. Finally, the $\indOutputs$-th output dimension of the $\indPoints$-th original map can be represented as:
\begin{equation}
\label{Eq:PCA_reconstruction}
    \outputPoint_{\indOutputs}^{(i)} \approx \dimensionMean_{\indOutputs} + \sum^{\nbasis}_{\indBasis=1} \coef_{\indBasis} (\inputPoint^{(i)}) \basis_{\indBasis,\indOutputs}
\end{equation}

All output dimensions for the $\indPoints$-th map can be computed directly in vector form as follows: 
\begin{equation}
\label{Eq:PCA_reconstruction_map}
    \outputPoint^{(\indPoints)} \approx \dimensionMean + \basisMatrixPCA^{\top} \coef(\inputPoint^{(\indPoints)})
\end{equation}

The PCA procedure was implemented in Python using the \textit{sklearn.decomposition} module of \textit{scikit-learn} library.

\subsubsection{Metamodeling by Gaussian process regression}
\label{subsubsec:GPR}

\newcommand{\meanGPR}{m}
\newcommand{\kernelGPR}{K}
\newcommand{\DoePoint}{\mathrm{x}}
%%%%%%%%% Gaussian process related macros

\renewcommand{\indGPR}{j}
\renewcommand{\GaussianProcess}{Z} 
\renewcommand{\MeanFunction}{m}
\renewcommand{\KernelFunction}{K}
\renewcommand{\condGP}{\GaussianProcess_n}
\renewcommand{\ConditionalMean}{\MeanFunction_c}
\renewcommand{\ConditionalKernel}{\KernelFunction_c}
\renewcommand{\condGPVector}{\GaussianProcess_n}
\renewcommand{\ObservationVector}{y^{\mathrm{obs}}}
\renewcommand{\KernelMatrix}{\KernelFunction(\DoE, \DoE)}
\renewcommand{\KernelVector}[1]{\KernelFunction(\DoE, #1)}
\renewcommand{\KernelVectorRow}[1]{\KernelFunction(#1, \DoE)}
\renewcommand{\NumberGPite}{N_{Z}}
\renewcommand{\nX}{N_X}
\newcommand{\deterministicModelGPR}{g}

Gaussian process regression (GPR) is a probabilistic metamodeling technique used to approximate the relationship between an input vector and a scalar model response. For more details, we refer to \cite{Williams1995}. Let $\deterministicModelGPR : \ \mathbb{X} \subseteq \R^\nvar \longrightarrow \R$ be a multivariate function, modeling a deterministic computer code. As a prior assumption, assume that $\deterministicModelGPR$ is one trajectory of a Gaussian process $Z$ on $\mathbb{X}$, denoted as $Z \sim \mathcal{GP}(\meanGPR,\kernelGPR)$, where $\meanGPR$ is the mean function and $\kernelGPR$ is the covariance function, or kernel. This prior information can be updated using a DoE $\DoE$ and respective observations $\ObservationVector_{\indPoints} = \deterministicModelGPR(\inputPoint^{(\indPoints)})$. This results in a posterior random process, that remains Gaussian. That conditional GP is denoted $\condGP \sim \mathcal{GP}(\meanGPR_c, \kernelGPR_c)$, where $\meanGPR_c$ and $\kernelGPR_c$ are given in closed-form by:
\begin{equation}
\begin{aligned}
\ConditionalMean(\inputPoint) &= \meanGPR(\inputPoint) + \KernelVectorRow{\inputPoint} \KernelMatrix^{-1} (\ObservationVector - \meanGPR(\DoE))  \qquad (\inputPoint \in \mathbb{X}) \\
\ConditionalKernel(\inputPoint, \inputPoint') &= \KernelFunction(\inputPoint, \inputPoint') - \KernelVectorRow{\inputPoint} \KernelFunction(\DoE, \DoE)^{-1} \KernelVector{\inputPoint'} \qquad (\inputPoint, \inputPoint' \in \mathbb{X})
\end{aligned}
\label{eq:kriging_mean_variance_intro}
\end{equation}
where $\ObservationVector$ is the vector of observation points, $\KernelMatrix = (\KernelFunction(\inputPoint^{(\indPoints)}, \inputPoint^{(j)}))_{1 \leq \indPoints,j \leq \nDoE}$ is the covariance matrix at design points, $\KernelVector{\inputPoint} = (\KernelFunction(\inputPoint^{(\indPoints)}, \inputPoint))_{1 \leq \indPoints \leq \nDoE}$ is the column vector of covariances between a new point $x$ and the design points, and $\KernelVectorRow{\inputPoint} = \KernelVector{\inputPoint}^\top$.

In the particular case of high-dimensional output prediction, the scalar response modeled by GPR corresponds to a basis coefficient, i.e $\deterministicModelGPR_\indBasis(\inputPoint) = \coef_{\indBasis}(\inputPoint)$ for each $\indBasis$-th retained basis component. The associated observation vector is the $\indBasis$-th column of the coefficient matrix: $\coefficientsMatrixPCA_{.,\indBasis} = (\coef_{\indBasis}(\inputPoint^{(\indPoints)}))_{\indPoints=1,\dots,\nDoE} \in \R^{\nDoE}$. One scalar GPR metamodel is then trained for each coefficient function $\inputPoint \mapsto \coef_\indBasis(\inputPoint)$, using the DoE $\DoE$ and the observations $\coefficientsMatrixPCA_{\cdot,\indBasis}$ as training data. Thus, a set of $\nbasis$ independent GPR metamodels is constructed: $Z_n^{(\indBasis)} \sim \mathcal{GP}(\ConditionalMean^{(\indBasis)}, \ConditionalKernel^{(\indBasis)})$, with $\indBasis=1,\dots, \nbasis$. Algorithm \ref{algo:prediction_intro} presents the PCA–GPR procedure, divided into an offline training phase and an online prediction phase. Both PCA and GPR procedures introduce errors which may propagate to the reconstructed maps; their explicit propagation is outside the scope of this study and remains a limitation of the present analysis.

Independent GPR metamodels were constructed for each retained PCA coefficient using the Python library \textit{GPflow} \citep{Matthews2017}. A zero mean function and a Matérn 5/2 covariance kernel with one length scale per input variable were adopted. For each metamodel, the kernel variance was initialized using the empirical variance of the corresponding PCA coefficient, while the length scales were initialized using the column-wise means of the training input sample. The kernel variance, length scales, and Gaussian-likelihood noise variance were then estimated by maximizing the marginal likelihood using the \textit{GPflow SciPy} optimizer iterations. Predictions at new input locations were obtained from the posterior mean of each GPR model.

\begin{algorithm}[H]
\caption{Offline construction and online prediction of a high-dimensional map using PCA and GPR.}
\label{algo:prediction_intro}
\begin{algorithmic}[1]

\STATE \textbf{Offline input:} DoE $\DoE$; training data $\dataMatrixPCA$; number of retained basis components $\nbasis$.

\STATE \textbf{Offline phase:}

\STATE Compute the dimension-wise mean $\dimensionMean$ with Eq. \ref{Eq:dimension_wise_mean} and center the training data, obtaining $\centeredData$.

\STATE Calculate the empirical covariance matrix $\covarianceMatrixPCA$ with Eq. \ref{Eq:covariance_matrix_PCA}.

\STATE Solve the eigenvalue problem
$\covarianceMatrixPCA \basis_{\indBasis}
= \lambda_{\indBasis}\basis_{\indBasis}$.

\STATE Sort the eigenvalues
$\lambda_1 \geq \dots \geq \lambda_{\nOutputDimensions}$
and the associated PCA basis components.

\STATE Retain the first $\nbasis$ PCA basis components and define the truncated PCA basis $\basisMatrixPCA$.

\STATE Project the centered data onto the PCA basis:
$\coefficientsMatrixPCA
= \centeredData \basisMatrixPCA^{\top}$.

\FOR{$\indBasis = 1$ \TO $\nbasis$}
    \STATE Define
    $\coefficientsMatrixPCA_{.,\indBasis}
    =
    \left(
    \coef_{\indBasis}
    \left(
    \inputPoint^{(\indPoints)}
    \right)
    \right)_{\indPoints=1,\dots,\nDoE}$.
    
    \STATE Train the $\indBasis$-th GPR metamodel
    $\GaussianProcess_{n}^{(\indBasis)}$
    using $\DoE$ and
    $\coefficientsMatrixPCA_{.,\indBasis}$.
\ENDFOR

\STATE \textbf{Online input:} New input scenario $\inputPoint^{*}$.

\STATE \textbf{Online phase:}

\FOR{$\indBasis = 1$ \TO $\nbasis$}
    \STATE Predict
    $\widehat{\coef}_{\indBasis}(\inputPoint^{*})$
    using
    $\GaussianProcess_{n}^{(\indBasis)}$.
\ENDFOR

\STATE Define the predicted basis-coefficient vector:
\[
\widehat{\coef}(\inputPoint^{*})
=
\left(
\widehat{\coef}_{1}(\inputPoint^{*}),
\dots,
\widehat{\coef}_{\nbasis}(\inputPoint^{*})
\right)^{\top}.
\]

\STATE Reconstruct the predicted high-dimensional map:
\[
\widehat{\outputPoint}(\inputPoint^{*})
=
\dimensionMean
+
\basisMatrixPCA^{\top}
\widehat{\coef}(\inputPoint^{*}).
\]

\STATE \textbf{return}
$\widehat{\outputPoint}(\inputPoint^{*})$.

\end{algorithmic}
\end{algorithm}

\subsubsection{Treatment of arrival times maps}
\label{subsubsec:treatment_arrival_times}

We note that the arrival-time data matrix contains NaN values at cells that were not flooded (see Eq. \ref{eq:max_flow_depth}). This prevents the direct application of PCA described in item \ref{subsubsec:PCA}. In this work, we address this issue using mean imputation, whereby each NaN value is replaced by the corresponding cellwise mean. GPR metamodeling is applied straightforwardly, as described in item \ref{subsubsec:GPR}, and new mean-imputed arrival times maps can be predicted for unexplored scenarios. Although this procedure preserves prediction accuracy for arrival times within the flooded region, it eliminates the flooded/non-flooded interface. To recover this interface, we use the predicted maximum flow depth maps $\widehat{H}_\ell(x^*)$ for the same unexplored scenario to define the flooded cells (if $\widehat{H}_\indOutputs(x^*) \geq \depthThresh, \ \indOutputs=1,\dots,\nOutputDimensions$) and non-flooded cells (otherwise). The non-flooded indices are removed from the corresponding predicted mean-imputed arrival times map, thereby reconstructing the flooded/non-flooded interface. Appendix \ref{appendix_arrival_times} illustrates the procedure using a scenario from the validation dataset.

\subsection{Global sensitivity analysis}
\label{subsec:GSA}

In this work, variance-based global sensitivity analysis (GSA) is employed. It relies on the Sobol-Hoeffding variance decomposition, where each normalised variance component defines a sensitivity index that measures the fraction of the total output variance attributable to individual inputs and their interactions. Let $\deterministicModel$ be any model that receives $\nvar$ independent input variables, whose vector is denoted by $\inputVectorRandom = (\inputPointRandom{1},\dots,\inputPointRandom{\nvar}) \in \R^{\nvar}$, with probability distribution $\probDist_\inputVectorRandom$. We denote the scalar model output as $\outputVectorRandom = \deterministicModel(\inputVectorRandom)$, which is assumed to be square-integrable and to have finite variance. For a square-integrable model with mutually independent inputs, the Sobol–Hoeffding decomposition uniquely expresses $\deterministicModel(\inputVectorRandom)$ as a sum of orthogonal component functions, each of which has zero mean with respect to its own variables. The variance decomposition formula can be written as follows:
\begin{equation}
\label{eq:variance_decomposition}
    \Var(\deterministicModel(\inputVectorRandom)) = \sum_{i=1}^{\nvar} \SobolUnnormalized{i} + \sum_{1\leq i <j \leq \nvar} \SobolUnnormalized{i,j} + \dots + 
    \SobolUnnormalized{1,\dots,\nvar}
\end{equation}

where $\SobolUnnormalized{\indexSet} = \Var(\deterministicModel_{\indexSet}(\inputVectorRandom_{\indexSet}))$ is the partial variance associated with the index set of input variables $\indexSet \subseteq \{ 1, \dots, \nvar \}$.

Each term of Eq. \ref{eq:variance_decomposition} can be interpreted as follows: $\SobolUnnormalized{i}$ refers to the contribution, or \textit{main effect}, of $\inputVectorRandom_i$ alone (unnormalized first-order Sobol' index), whereas $\SobolUnnormalized{i,j}$ quantifies the interaction effect between $\inputVectorRandom_i$ and $\inputVectorRandom_j$ (unnormalized second-order Sobol' index). This interpretation and terminology extend to higher-order terms. Two types of sensitivity indices are used in this work: the first-order and the total Sobol' index. The total Sobol' index of a given variable refers to its main effect and all interactions involving this variable. In practice, except for some classes of models, e.g. polynomial chaos expansion \citep{Sudret2008}, Sobol' indices are estimated by empirical methods, more specifically through \textit{pick-freeze} (PF) schemes \citep{Saltelli2008, DaVeiga2021}. Estimated quantities are denoted by $\widehat{(\cdot)}^{pf}$ in this work.

In the context of high-dimensional outputs, sensitivity indices are represented by sensitivity maps (SMs) \citep{Marrel2011, Sao2025}, where a Sobol' index is defined at each output dimension for each input variable. Furthermore, concise sensitivity indices, known as generalized sensitivity indices (GSIs), can be used to quantify the variance-weighted average of sensitivity indices across all output dimensions for a given input variable \citep{Lamboni2011, Perrin2021, Sao2025}. Within the reduced-order framework, two approaches are considered: a \textit{basis-derived} approach and a \textit{dimension-wise} approach. For a complete description of the methods, we refer to \citep{Li2020, Sao2025} for the basis-derived approach and \citep{Marrel2011} for the dimensionwise approach. The basis-derived approach computes the covariance matrices of reduced-basis coefficients through a scalar-valued PF scheme. Then, the basis eigenvectors are used to project the covariance matrices into the original output space, resulting in the Sobol' index. This approach is computationally more efficient, as demonstrated by \citep{Sao2025}. The basis-derived closed-form formula for the Sobol' index of the $\indOutputs$-th output dimension $\pfEstim{\sobolClosed{\indexSet}}(\outputHighDimensional) \in \R$ is written as:
\begin{equation}
\label{eq:GSA_reprojection}
    \pfEstim{\sobolClosed{\indexSet}}(\outputHighDimensional) = \frac{\basisVec^{\top} \ \pfEstim{\sobolClosedD{\indexSet}}(\coefPred) \ \basisVec}{\basisVec^{\top} \ \pfEstim{\totalVariance}(\coefPred) \ \basisVec}
\end{equation}

where $\pfEstim{\sobolClosedD{\indexSet}}(\coefPred) \in \R^{\nbasis \times \nbasis}$ is the covariance matrix of predicted basis coefficients related to $\indexSet$ and $\pfEstim{\totalVariance}(\coefPred) \in \R^{\nbasis \times \nbasis}$ is the overall covariance matrix of the sampled predicted basis coefficients. The basis-derived $ \pfEstim{\GSI_\indexSet} \in \R$ is defined as follows:
\begin{equation}
\label{eq:GSA_GSI}
\pfEstim{\GSI_\indexSet} = \frac{\Tr{\pfEstim{\sobolClosedD{\indexSet}}(\coefPred) \, \GramMatrix}}{\Tr{\pfEstim{\totalVariance} (\coefPred)\, \GramMatrix}}
\end{equation}

The dimension-wise approach estimates the Sobol' index directly in the original output space through a scalar-valued PF scheme. For the $\indOutputs$-th output dimension, the Sobol' index related to $\indexSet$ is written as: 
\begin{equation}
\label{eq:sobol_scalar_valued}
    \pfEstim{\sobolClosed{\indexSet}}(\outputHighDimensional) = \frac{\pfEstim{\sobolClosedD{\indexSet}}(\outputHighDimensional)}{\pfEstim{\totalVariance}(\outputHighDimensional)}
\end{equation}

where, for the $\indOutputs$-th output dimension, $\pfEstim{\sobolClosedD{\indexSet}}(\outputHighDimensional) \in \R$ is the unnormalized Sobol' index of $\indexSet$ and $\pfEstim{\totalVariance}(\outputHighDimensional) \in \R$ is the overall variance. The dimensionwise $\pfEstim{\GSI_\indexSet} \in \R$ is defined as follows:
\begin{equation}
\label{eq:gsi_estimation}
\pfEstim{\GSI}_{\indexSet} = \frac{\sum_{\indOutputs=1}^{\nOutputDimensions} \pfEstim{\totalVariance}_\indOutputs \ \pfEstim{\sobolClosed{\indexSet}}(\outputHighDimensional) }{\sum_{\indOutputs=1}^{\nOutputDimensions} \pfEstim{\totalVariance}_\indOutputs }
\end{equation}

In order to estimate SMs of maximum flow depth, the basis-derived approach is employed. In the case of arrival times maps, basis coefficients no longer represent the final maps due to the specific pre- and post-treatments (see item \ref{subsubsec:treatment_arrival_times}). Therefore, the dimension-wise approach is employed instead.

\section{Results and discussion}
\label{sec:results}

The ICOLD test case was simulated using HEC-RAS, and the resulting data were analyzed using the probabilistic framework. In total, $500$ scenarios were simulated, with $250$ scenarios allocated to each of the training and validation sets. Negligible difference in sensitivity indices was observed by incrementing the training set to size $\nDoE = 450$. Concerning dimensionality reduction, $\nbasis = 15$ components ($99\%$ explained variance) and $\nbasis = 40$ components ($95\%$ explained variance) were used to represent the maximum flow depth and arrival times maps, respectively. For sensitivity analysis, the Janon-Monod \citep{Janon2014} and Jansen \citep{Jansen1999} estimators are employed to compute the first-order and total Sobol' indices, respectively. We adopted samples of size $\nPF = 10000$ and $\nPF = 5000$ to compute maximum flow depth and arrival times sensitivity indices, respectively. The bootstrap technique was used to evaluate the PF estimation error. For each of the 100 bootstrap replicates, the $\nPF$ rows of the PF samples were resampled with replacement, using the same resampled indices for all coupled sample matrices to preserve the PF structure. The first-order and total Sobol’ indices were then recomputed from each resampled dataset. The median and confidence bounds therefore quantify the variability of the PF estimator conditional on the trained metamodels. Therefore, these confidence bounds do not account for PCA truncation error or GPR prediction error; these errors are further discussed in \citep{Legratiet2014, Li2020} The validation of these choices is presented in Appendix \ref{app:validation}.

\subsection{Global sensitivity analysis of maps}
\label{subsec:GSA_results}

This subsection presents GSA results of maximum flow depth (item \ref{subsubsec:maxdepth}) and arrival times (item \ref{subsubsec:arrival_times}). Concise overviews of sensitivity are given by first-order and total GSI and, then, spatially-varied descriptions of the Sobol' indices are given by SMs. 

\subsubsection{Maximum flow depth}
\label{subsubsec:maxdepth}

\begin{figure}[H]
    \centering
    \includegraphics[width=0.95\linewidth]{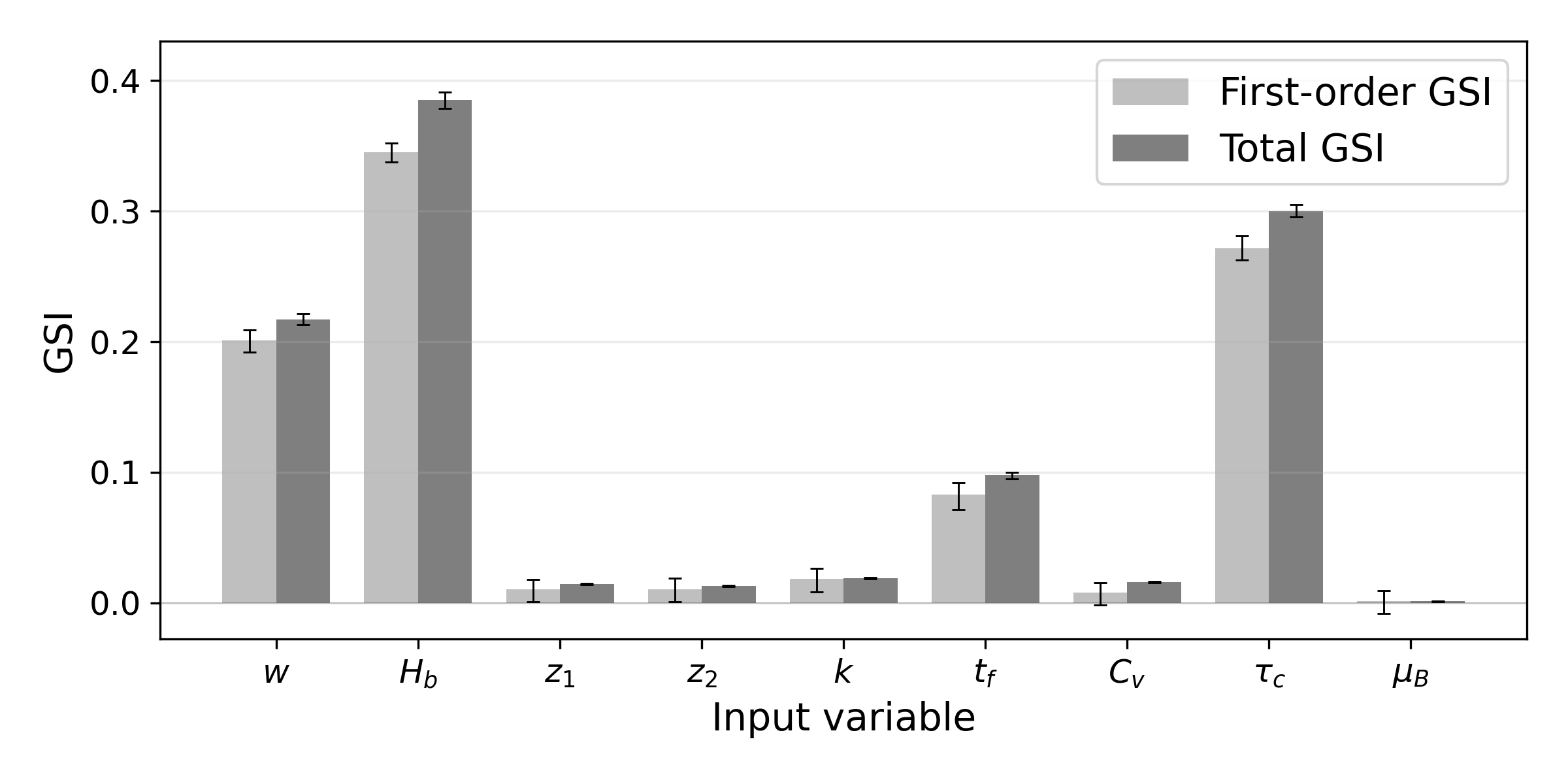}
    \caption{Maximum flow depth: first-order and total generalized sensitivity indices. Error bars were calculated with $100$ bootstrap repetitions.}
    \label{fig:GSI_maxdepth}
\end{figure}

Figure \ref{fig:GSI_maxdepth} shows the first-order and total GSI for maximum flow depth. It indicates that maximum flow depth is mainly controlled by some of the breach parameters ($\bottomElevation, \bottomWidth, \breachTime$) and viscoplasticity of tailings ($\yieldStress$). Other breach parameters ($\rightSlope, \leftSlope, \breachCoeff$) and properties of tailings ($\volumetricConc, \plasticViscosity$) are negligible. The relatively small differences between first-order and total indices indicate limited interaction effects.

Concerning the breach parameters, $\bottomElevation$ determines the elevation of the breach, which controls the released volume by the upstream reservoir. Released volume is known to significantly affect the downstream characteristics of the flow, such as runout and flow depths \citep{Rico2008a}. Other breach parameters ($\bottomWidth, \rightSlope, \leftSlope$) determine the breached area, which affect the released hydrograph. Among those, $\bottomWidth$ is the parameter that most strongly affects the breached area and, consequently, the maximum flow depth, according to Fig.~\ref{fig:GSI_maxdepth}. The influence of $\bottomElevation$ and $\bottomWidth$ on maximum flow depth was also reported by \citep{kalinina2020} in a study of water-retaining dam-breach flow. The temporal development of the breach is governed by $\breachTime$ and has a secondary influence on maximum flow depth. 

Concerning tailings properties, $\yieldStress$ represents the dominant resistance mechanism, as it directly controls the basal shear stress (Eq.~\ref{eq:basal_shear_stress}). The plastic viscosity $\plasticViscosity$ scales with the shear rate and exhibits a minor influence because shear rates remain low over most of the flow domain. Volumetric concentration $\volumetricConc$ mainly acts as a scaling factor for the mixture density $\densityMixture$ (Eq.~\ref{eq:density_mixture}) and does not control any dominant term in the shallow-water equations; thus, it is negligible. For other modeling approaches, however, $\volumetricConc$ may play a more significant role if rheological parameters are explicitly defined as functions of concentration.

\begin{figure}[H]
    \centering
    \includegraphics[width=0.95\linewidth]{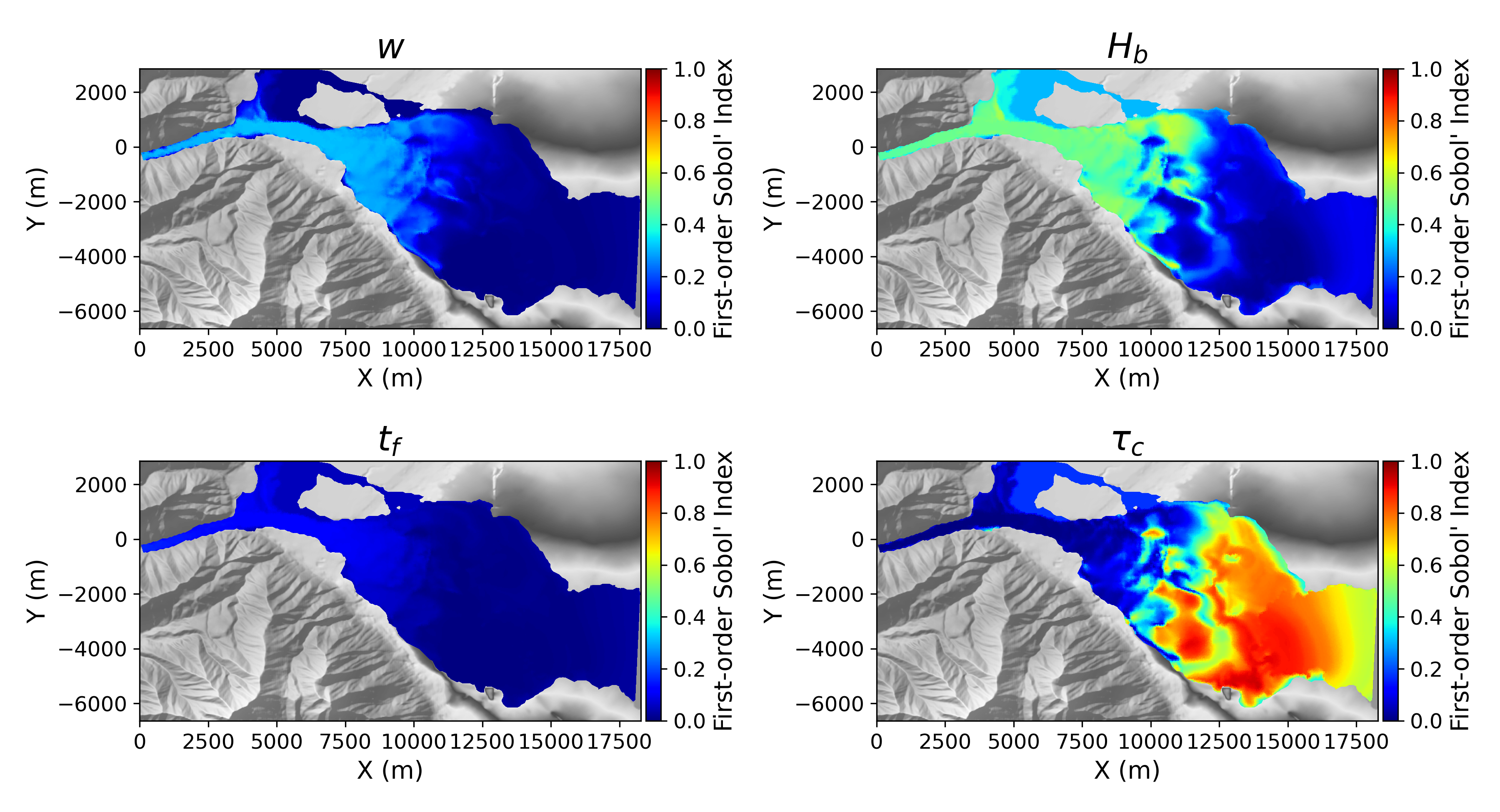}
    \caption{Maximum flow depth: first-order sensitivity maps for four of the most influential input variables. Values represent the median of 100 bootstrap replicates. Warm colors indicate greater sensitivity to the corresponding input variable.}
    \label{fig:sensitivity_map_maxdepth}
\end{figure}

In order to detail the observations obtained by the GSI analyses, Fig. \ref{fig:sensitivity_map_maxdepth} shows the spatially-varied first-order Sobol' indices of the four most important variables $\bottomElevation, \bottomWidth, \breachTime, \yieldStress$. A strong spatial variability of Sobol' indices is observed, demonstrating the coexistence of different physical mechanisms along the flow path. 

By inspecting the SMs of $\bottomWidth, \bottomElevation, \breachTime$, a region containing non-negligible Sobol' indices can be identified in the upstream and central parts of the flooded area. It indicates that the propagation of volume through the boundary condition significantly affects maximum flow depth in that region, consistently with an inertia-dominated regime. On the other hand, breach parameters have negligible Sobol’ indices farther downstream, indicating little influence on maximum flow depth in these areas.

The SM for $\yieldStress$ shows an area presenting significant Sobol' indices in the central and downstream parts of the domain. Farther downstream, $\yieldStress$ completely dominates, since those are reached only by scenarios with sufficiently low yield stress, such that variability in maximum flow depth is almost entirely driven by $\yieldStress$. Physically, this corresponds to a viscoplastic flow regime in which basal shear stress controls energy dissipation and inertia induced by the initial breach becomes negligible. A region where significant Sobol' indices of both $\bottomElevation$ and $\yieldStress$ coexist can be observed, marking the transition between the breach- and $\yieldStress$-dominated regions. Overall, Fig.~\ref{fig:sensitivity_map_maxdepth} complements the global GSI analysis by revealing that breach parameters govern maximum flow depth near the dam, whereas rheological resistance, represented by $\yieldStress$, dominates far downstream.

\subsubsection{Arrival times}
\label{subsubsec:arrival_times}

Because arrival time is undefined in dry cells, Sobol’ indices were computed only for cells flooded in all 5000 realizations of the corresponding dimension-wise PF samples. Consequently, the arrival-time sensitivity maps are restricted to the always-flooded portion of the domain. In the remaining cells, evaluating variability of arrival times would require conditioning on the event that the cell is flooded. Such conditioning changes the input distribution and may induce statistical dependence among inputs that are independent in the original probabilistic model, because only combinations of parameters leading to flooding are retained. Therefore, standard Sobol’ indices cannot be applied directly in these regions. Accordingly, the results presented here quantify the influence of the uncertain inputs on arrival time variability only within the always-flooded domain.

\begin{figure}[H]
    \centering
    \includegraphics[width=0.95\linewidth]{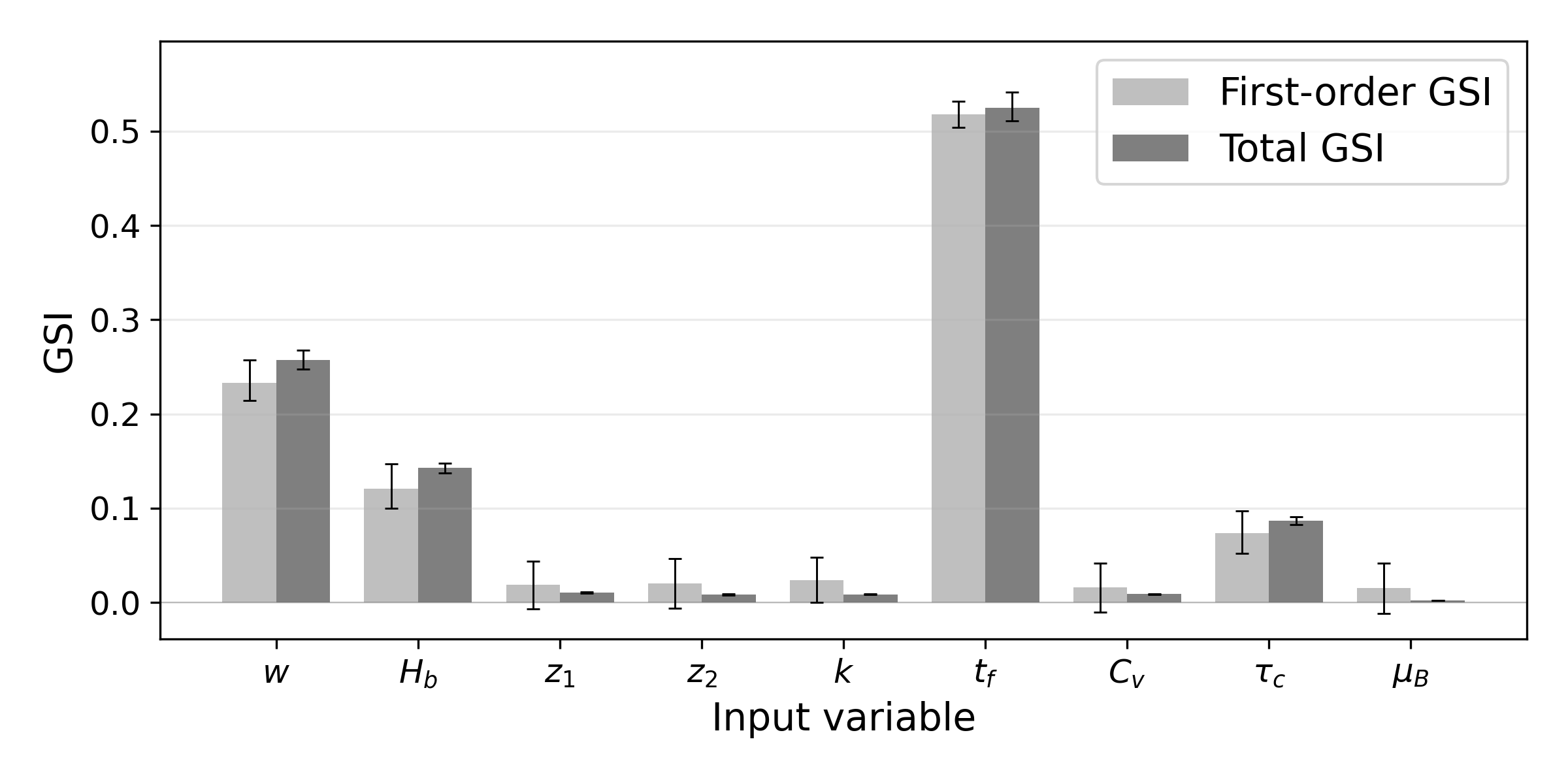}
    \caption{Arrival times: first-order and total generalized sensitivity indices. Error bars were calculated with $100$ bootstrap repetitions.}
    \label{fig:GSI_arrival}
\end{figure}

Figure~\ref{fig:GSI_arrival} presents the first-order and total generalized sensitivity indices for arrival times. Similarly to the maximum flow depth case, arrival times are controlled by breach parameters ($\bottomElevation, \bottomWidth, \breachTime$) and $\yieldStress$. However, there are some differences in this case. First, $\breachTime$ is the most influential parameter, since it controls the release of volume and, consequently, how quickly the tailings flow reaches a certain location. Second, first-order and total indices have negligible differences, indicating negligible interaction effects among variables. Third, $\yieldStress$ is non-negligible but plays a secondary role. These observations indicate that uncertainty in arrival times is mainly driven by the dynamics of breach development, rather than by rheological resistance alone.

\begin{figure}[H]
    \centering
    \includegraphics[width=0.95\linewidth]{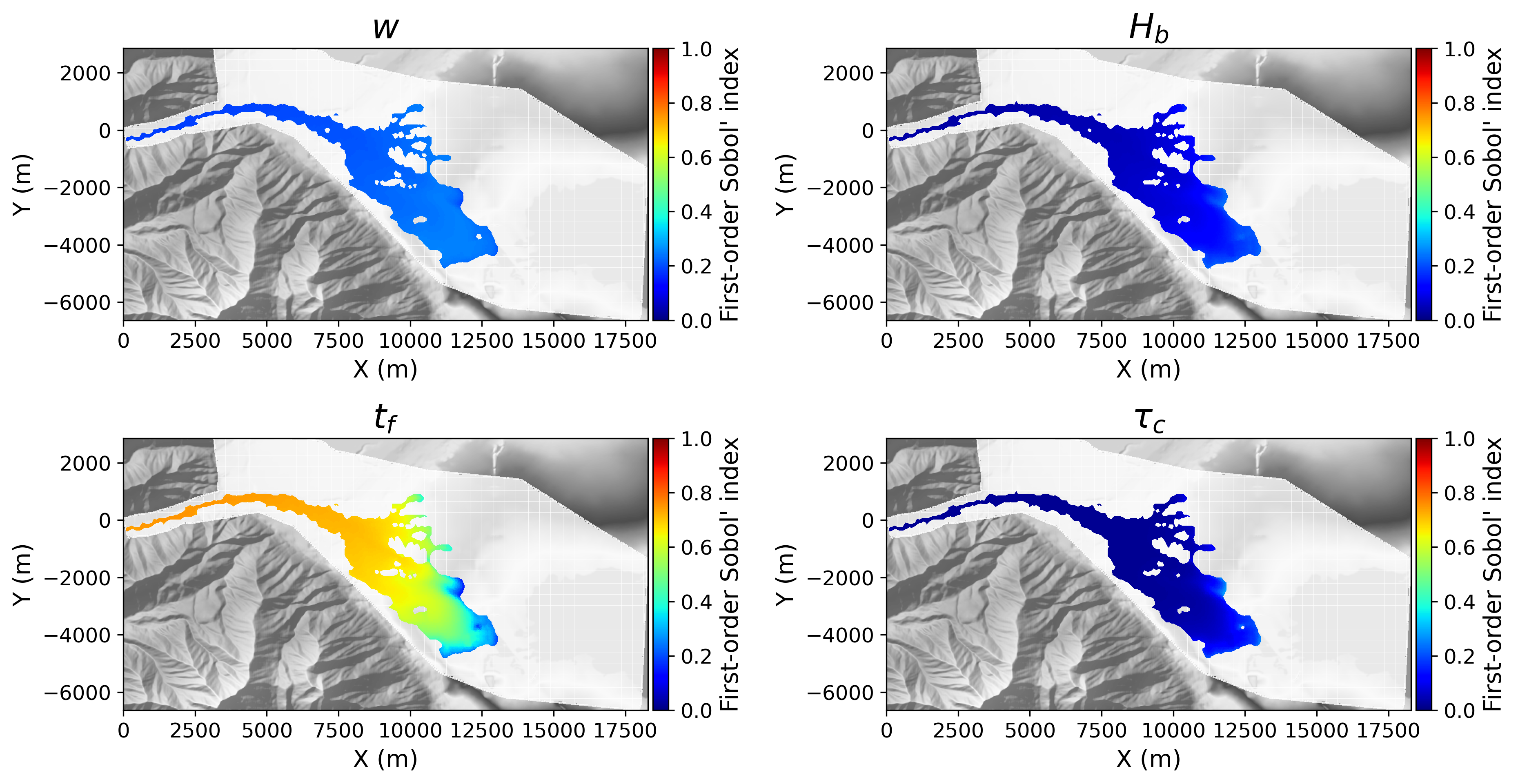}
    \caption{Arrival times: first-order sensitivity maps of four of the most influential input variables. White region corresponds to cells without a defined Sobol' index. Values correspond to the median of $100$ bootstrap repetitions. Hot colors indicate higher sensitivity to the corresponding input variable.}
    \label{fig:sensitivity_map_arrival}
\end{figure}

Figure \ref{fig:sensitivity_map_arrival} shows the SMs related to arrival times maps for the most influential variables. The SMs confirm that $\breachTime$ is the dominant control on arrival time variability within the flooded domain, indicating it primarily controls the arrival times at upstream and central areas. Farther downstream, $\breachTime$ loses gradually its importance, shifting the influence to $\bottomWidth, \bottomElevation, \yieldStress$.

\subsection{Probabilistic analyses of flooding outputs}
\label{subsec:probabilistic_analyses}

In the context of tailings dam-breach analyses, probabilistic outputs provide information that is directly relevant for hazard assessment and decision-making. Algorithm \ref{algo:prediction_intro} was used to predict $10000$ unexplored scenarios of maximum flow depth and arrival time. These scenarios allowed the determination of maps of flooding probability (item \ref{subsubsec:flooding_probability}); furthermore, distributions of maximum flow depth and arrival times at control sections and associated Sobol' indices were evaluated at item \ref{subsubsec:maximum_flow_depth_prob}.

\subsubsection{Flooding probability}
\label{subsubsec:flooding_probability}

\begin{figure}[H]
    \centering
    \includegraphics[width=0.95\linewidth]{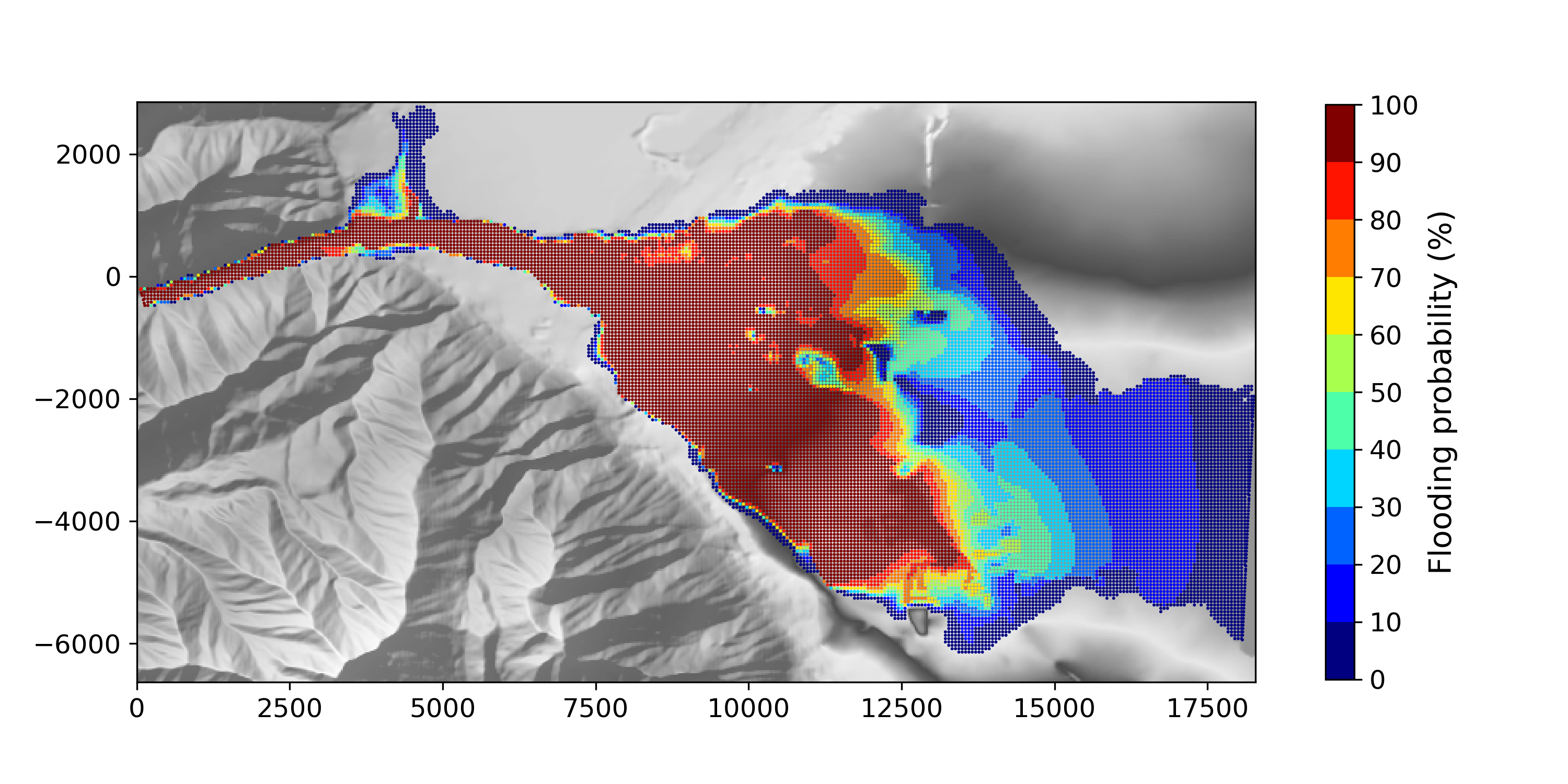}
    \caption{Map of flooding probability generated with $10000$ predicted scenarios. Hot colors correspond to higher probability of flooding.}
    \label{fig:flooding_probability}
\end{figure}

Flooding probability $\floodProb{}$ is computed from the set of maximum flow depth maps by defining a binary flooding criterion. A cell is considered flooded if $\maxFlowDepth_{\indOutputs} > \depthThresh$, where the threshold $\depthThresh = 0.31~\mathrm{m}$ is commonly adopted in flood-risk studies as a reference threshold value. For a given cell $\indOutputs$, the flooding probability is then defined as
\begin{equation}
\label{eq:flooding_probability}
\floodProb{\indOutputs} = \frac{1}{\nPF} \sum_{s=1}^{\nPF} \mathbf{1}\left\{ \maxFlowDepth^{(s)}_{\indOutputs} > \depthThresh \right\},
\end{equation}
where $\mathbf{1}\{\cdot\}$ denotes the indicator function, which equals 1 when the condition is satisfied and 0 otherwise. The resulting flooding probability map for $10000$ unexplored maps is shown in Fig.~\ref{fig:flooding_probability}.

Figure~\ref{fig:flooding_probability} reveals three distinct spatial regimes. Over the first approximately $12~\mathrm{km}$ downstream of the dam, encompassing the V-shaped valley and a substantial portion of the flat region, flooding probability is very high ($\floodProb{} > 80\%$). In this zone, nearly all simulated scenarios exceed the threshold $\depthThresh$, indicating that the binary flooding classification shows limited sensitivity with respect to input uncertainty. This behavior reflects the strong confinement imposed by the topography, specially at $x\sim 4.5~\mathrm{km}$ where a natural levee along the left margin confines the flow to the upper part of the region. Therefore, topographic confinement leads to systematically large flow depths regardless of parameter variability.

Farther downstream, a transition region between $x \sim [12.5, 16] \ \mathrm{km}$ is observed, characterized by intermediate flooding probabilities ($20\% < \floodProb{} < 80\%$). In this region, flooding probability is sensitive to input uncertainty, as relatively small variations in the governing parameters produce significant changes in flood extent. This zone therefore represents a critical area from a hazard-assessment perspective, where uncertainty in model inputs translates directly into uncertainty in flood delineation. Finally, regions with low flooding probability ($\floodProb{} < 20\%$) are located far from the dam and laterally with respect to the main flow path. Flooding in these areas occurs only for a limited subset of scenarios, primarily those associated with low yield stress, in agreement with the sensitivity patterns identified for maximum flow depth in Section~\ref{subsubsec:maxdepth}.

\subsubsection{Distributions of maximum flow depth and arrival times}
\label{subsubsec:maximum_flow_depth_prob}

Beyond flood extent, it is of practical interest to characterize the probabilistic distributions of maximum flow depth and arrival times. To facilitate visualization and interpretation, the domain is divided into $12$ cross-sections at fixed $\text{x}$-coordinates, starting at $x = 0.585 \ \mathrm{km}$ from the dam and spaced at $1 \ \mathrm{km}$ intervals, as indicated in Fig.~\ref{fig:hypothetical_dam_terrain}. Let $\maxFlowDepth=\maxFlowDepth(\text{x}, \text{y})$ and $\arrivalTime=\arrivalTime(\text{x}, \text{y})$, where $(\text{x}, \text{y})$ is the spatial coordinate of a given cell. For each cross-section, two scalar quantities are extracted from the predicted maps: the cross-sectional maximum flow depth and the minimum arrival time:
\begin{equation}
\label{eq:maximum_maxdepth}
\maxFlowDepth^{\mathrm{sec}}(\text{x}) = \max_{\text{y}} \maxFlowDepth(\text{x},\text{y}) \quad ; \quad \arrivalTime^{\mathrm{sec}}(\text{x}) = \min_{\text{y}} \arrivalTime(\text{x},\text{y})
\end{equation}

\begin{figure}[H]
    \centering
    \includegraphics[width=0.95\linewidth]{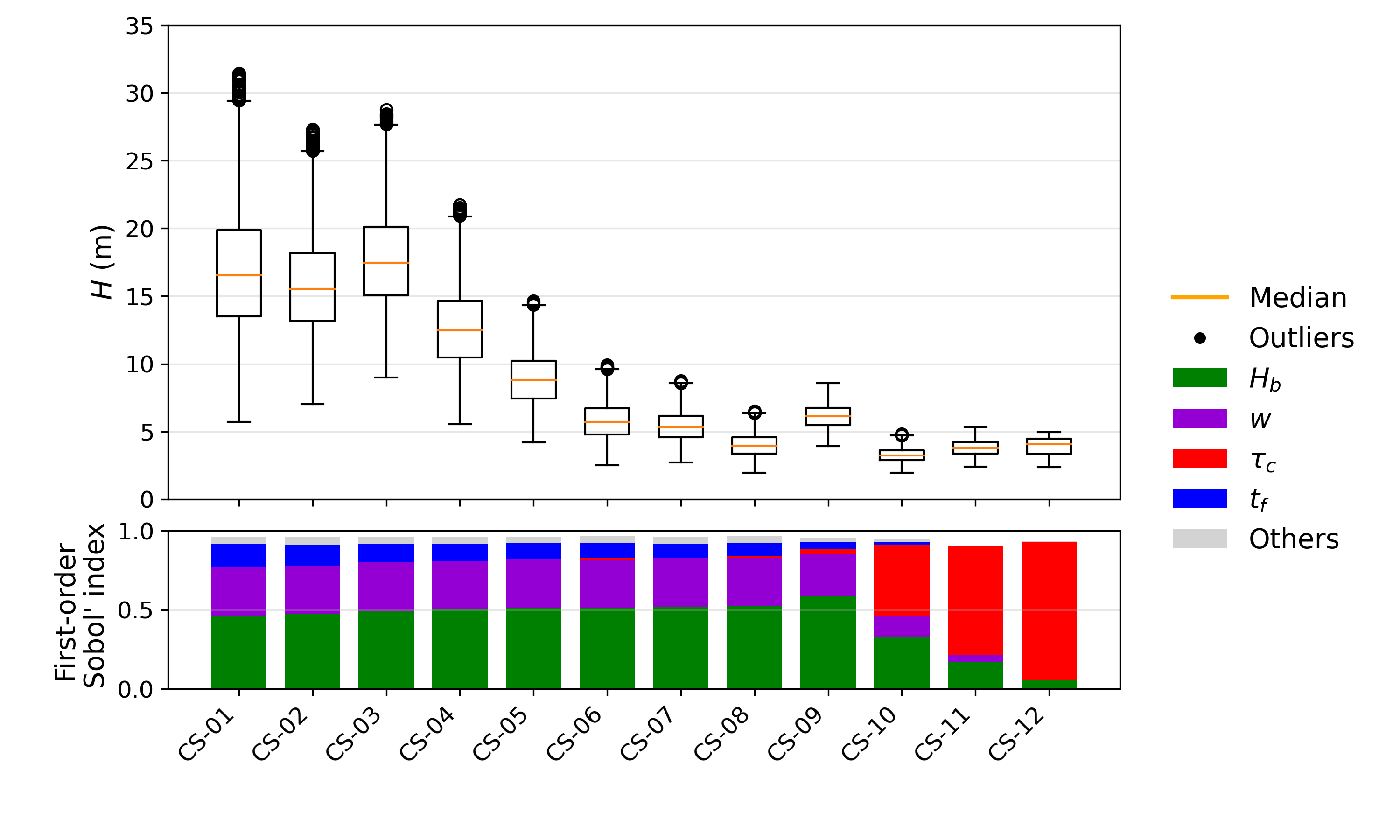}
    \caption{Maximum flow depth: top panel presents data distribution of $\maxFlowDepth^{sec}(\text{x})$ at each cross-section; bottom panel shows first-order Sobol' indices of the most influential variables over $\maxFlowDepth^{sec}(\text{x})$.}
    \label{fig:maximum_flow_depth_boxplots}
\end{figure}

\begin{figure}[H]
    \centering
    \includegraphics[width=0.95\linewidth]{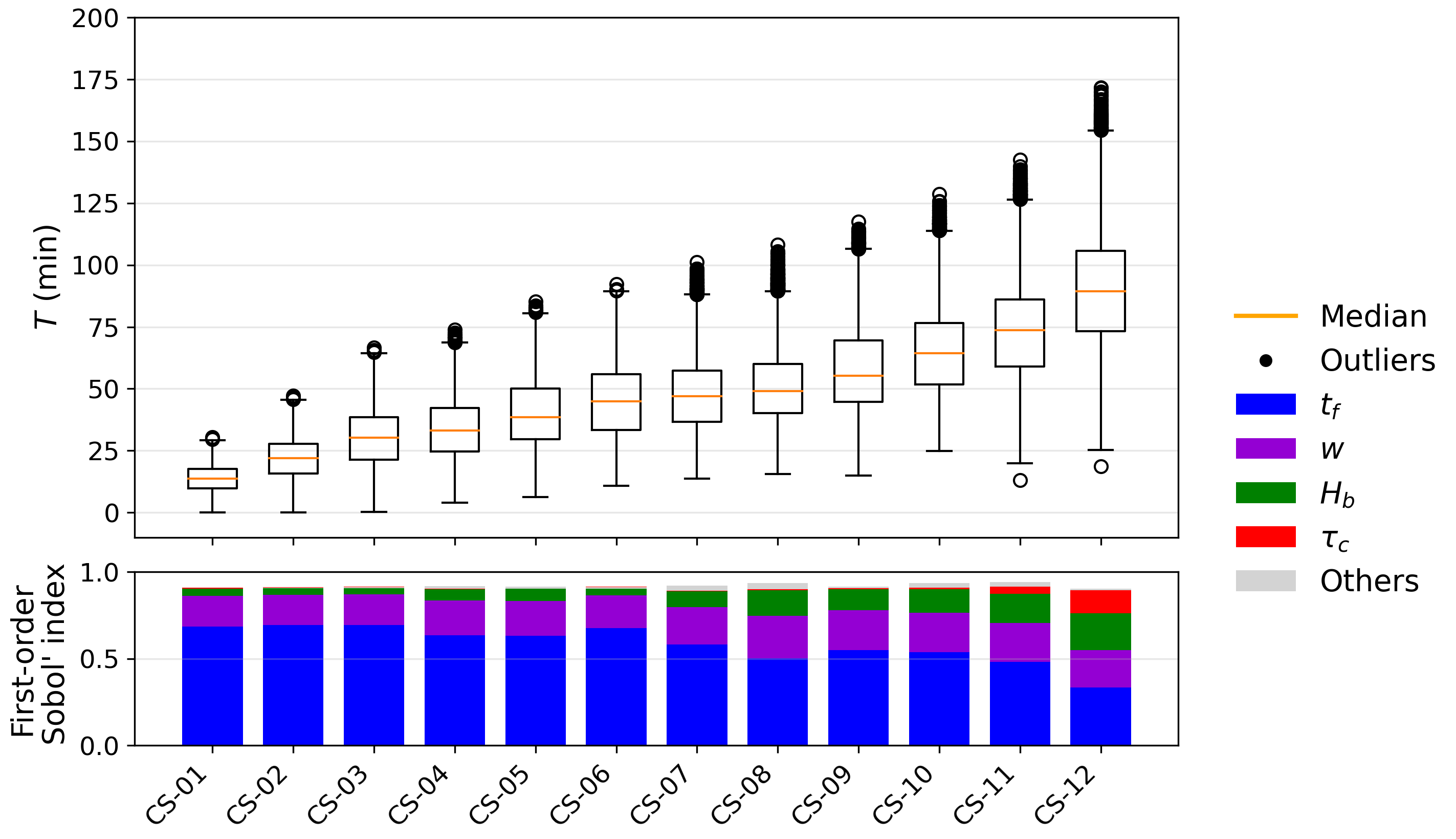}
    \caption{Arrival times: top panel presents data distribution of $\arrivalTime^{sec}(x)$ at each cross-section; bottom panel shows first-order Sobol' indices of the most influential variables over $\arrivalTime^{sec}(x)$.}
    \label{fig:arrival_boxplots}
\end{figure}

These quantities represent conservative metrics commonly used in risk-oriented analyses, as they capture the maximum depth and the earliest wavefront arrival at each cross-section. For each section, this procedure yields $10000$ samples of $\maxFlowDepth^{\mathrm{sec}}$ and $\arrivalTime^{\mathrm{sec}}$, which are summarized using boxplots, which present the median (50th percentile), interquartile ranges (25th and 75th percentiles), whiskers and outliers. In addition, first-order Sobol' indices are computed for each cross-section using a scalar-valued PF scheme with PF sample size of $\nPF=10000$. The most influential input variables are displayed in the lower panels of Figs.~\ref{fig:maximum_flow_depth_boxplots} and \ref{fig:arrival_boxplots}.

Figure~\ref{fig:maximum_flow_depth_boxplots} shows the distributions of $\maxFlowDepth^{\mathrm{sec}}$ along the downstream direction. Median values decrease monotonically from approximately $16$--$18~\mathrm{m}$ near the dam (CS-01, CS-02 and CS-03) to about $3$--$5~\mathrm{m}$ at the farthest cross-sections (from CS-06 onward). This trend is physically expected, as the flow is initially confined within the narrow valley and subsequently spreads over the flat region, leading to reduced maximum depths. The interquartile range also narrows downstream, indicating a progressive reduction of uncertainty in $\maxFlowDepth^{\mathrm{sec}}$. Consistent with the sensitivity analysis of Section~\ref{subsubsec:maxdepth}, Sobol' indices show that breach-related parameters dominate uncertainty near the dam, whereas yield stress becomes dominant farther downstream. However, the results show that the uncertainty in the maximum flow depth is very sensitive to breach parameters ($\bottomWidth, \bottomElevation, \breachTime$), with substantial variability across the first nine cross-sections. On the other hand, for the last three cross-sections, $\yieldStress$ causes comparatively less variation in the maximum flow depth. Therefore, even though yield stress presents high Sobol' indices farther downstream from the dam, it does not necessarily imply large variations in maximum flow depth.

In contrast, Fig.~\ref{fig:arrival_boxplots} indicates that the dispersion of $\arrivalTime^{\mathrm{sec}}$ increases with distance from the dam. Scenarios characterized by low viscoplastic resistance, rapid breach development, and larger breach cross-sectional areas populate the lower tail of the distributions, corresponding to earlier arrivals. The associated Sobol' indices reveal that breach-related parameters, particularly $\breachTime$ and $\bottomWidth$, remain influential across all cross-sections, while $\yieldStress$ plays a secondary role in the minimum arrival times. This behavior is consistent with the definition of $\arrivalTime^{\mathrm{sec}}$ as the earliest arrival within each cross-section, which emphasizes early-time propagation dynamics controlled by breach development rather than by resistance mechanisms governing flow persistence. 

Overall, these probabilistic analyses highlight the complementary nature of different flood metrics. Flooding probability rapidly saturates in confined regions and is most informative near the flood-extent boundaries. Maximum flow depth exhibits decreasing uncertainty downstream, whereas arrival times show increasing dispersion driven by early-wave propagation mechanisms. Together, these results illustrate how uncertainty propagates differently across flood characteristics and emphasize the value of combining probabilistic outputs with sensitivity analysis for comprehensive TDBA assessments.

\section{Conclusions and further work}
\label{sec:conclusion}
This study applied an efficient probabilistic methodology combining uncertainty propagation and global sensitivity analysis to tailings dam-breach flows, while addressing methodological challenges associated with computationally expensive simulations and high-dimensional outputs. The framework produced flooding-probability maps, output distributions, and both aggregated and spatially distributed sensitivity indices. These outputs supported both the quantification and interpretation of output uncertainty, as well as the physical interpretation of the simulated flow behaviour. A benchmark test case was explored and deterministic simulations were performed by using HEC-RAS v6.6. The two main contributions of this work are complementary:

\begin{enumerate}
    \item The proposed framework provides fast and accurate predictions of output maps, allowing the computation of sensitivity maps, which are useful for physical interpretation. These maps reveal strong spatial heterogeneity of Sobol' indices and, consequently, show different physical mechanisms in different points of the terrain, such as areas dominated by breach parameters and areas dominated by viscoplasticity. These spatially-varied mechanisms cannot be captured by global indicators, e.g. aggregated sensitivity indices.

    \item From an applied perspective, the framework enables statistical analysis of outputs relevant to hazard and risk assessment. The resulting distributions quantify uncertainty and provide information based on arrival times (early-arrival zones and warning-time assessment), on maximum flow depth (spatial variability in flood severity) and on flooding probability.
\end{enumerate}

Specifically for the ICOLD test case, sensitivity maps showed that breach-related parameters dominate the region near the dam, whereas viscoplastic resistance represented by $\yieldStress$ are dominant farther downstream, allowing the identification of two physically distinct flow regimes. The near-dam region is primarily controlled by breach development and release dynamics, while downstream regions are governed by viscoplastic effects. In addition, probabilistic analyses provided flooding probability maps and cross-sectional distributions of maximum flow depth and arrival times, enabling the assessment of many possible scenarios. Estimating sensitivity indices at each cross-section helped identify the sources of output uncertainty and confirmed the spatial variability observed in the sensitivity maps.

Future work could improve the estimation of errors associated to PCA and GPR modeling \citep{Legratiet2014, Li2020, Sao2025b}, as well as the treatment of arrival times by using a more sophisticated imputation technique to perform dimensionality reduction with PCA, e.g. weighted-PCA \citep{Chaleshtori2024}, or by using a non-linear basis expansion technique, such as autoencoders \citep{Rohmer2023}. In terms of probabilistic modeling, dependencies among input variables should also be explored \citep{DaVeiga2015}, especially in the context of rheological modelling by linking the volumetric concentration with rheological parameters \citep{Obrien1988}. Regarding future applications, the methodology could be applied to cases involving quasi-instantaneous failures, such as the Brumadinho disaster, or gradual breaches, such as the Mariana disaster. It could also be extended to more decision-relevant outputs, including damage and loss-of-life estimates \citep{Lumbroso2021, Silva2023}. Despite these limitations, this work ultimately contributes to advancing the study of tailings dam-breach flows by proposing a modular, non-intrusive and general probabilistic framework, that can be extended to other deterministic models, dimensionality-reduction methods, metamodeling techniques and hazard-related outputs.

\appendix
\section{Treatment of arrival times maps}
\label{appendix_arrival_times}

\begin{figure}[H]
    \centering
    \includegraphics[width=0.95\linewidth]{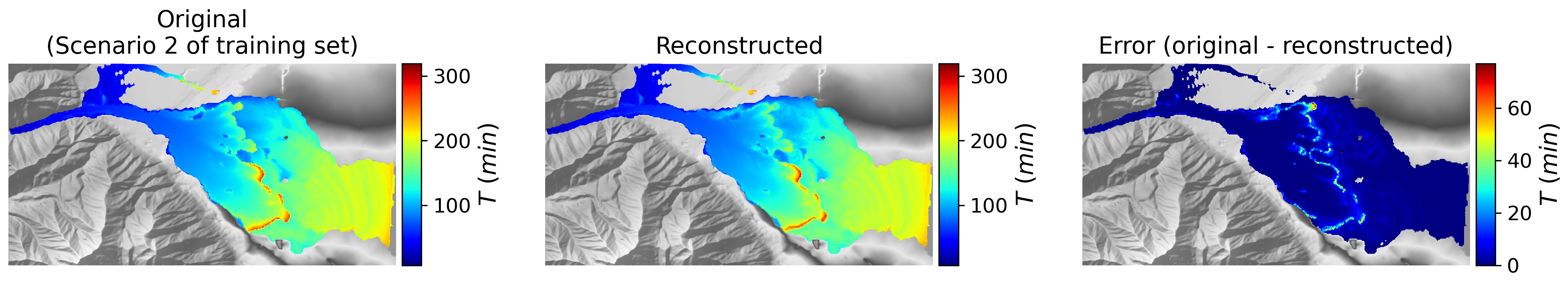}
    \caption{Comparison of original mean-imputed map with a PCA-reconstructed map.}
    \label{fig:app_mean_imputation}
\end{figure}

Dimensionality reduction and metamodeling of arrival times maps cannot be applied straightforwardly because of NaN values within the data-matrix (see item \ref{subsubsec:treatment_arrival_times}). This section illustrates the procedure of mean-imputation and masking of flooded/non-flooded cells. Figure \ref{fig:app_mean_imputation} shows the PCA reconstruction of mean-imputed maps and the respective error. 

\begin{figure}[H]
    \centering
    \includegraphics[width=0.95\linewidth]{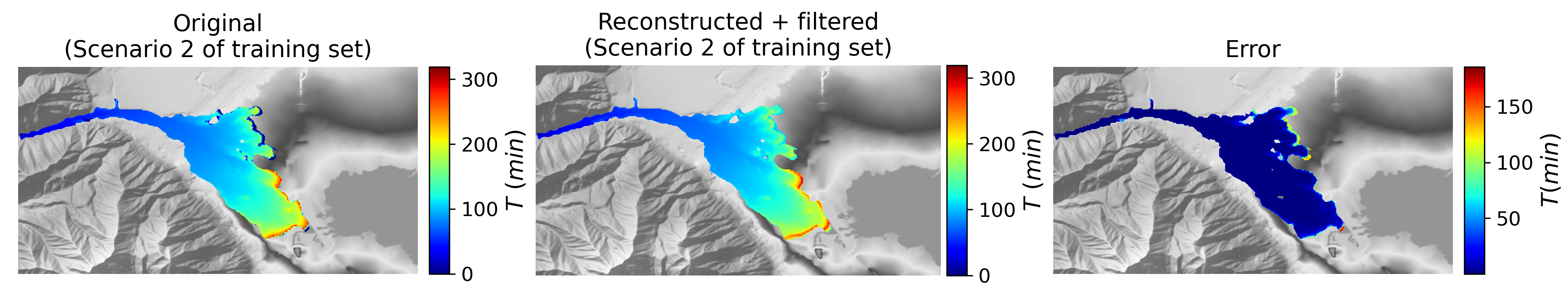}
    \caption{Comparison of filtered original map with filtered PCA-reconstructed map.}
    \label{fig:app_reconstruction_cropped}
\end{figure}

Next, to recover the flooded/non-flooded interface, the map of maximum flow depth corresponding to the same input scenario is used. Flooded cells ($h > 0.31 \text{m}$) are kept and non-flooded cells are discarded. This step is demonstrated in Fig. \ref{fig:app_reconstruction_cropped}.

\begin{figure}[H]
    \centering
    \includegraphics[width=0.95\linewidth]{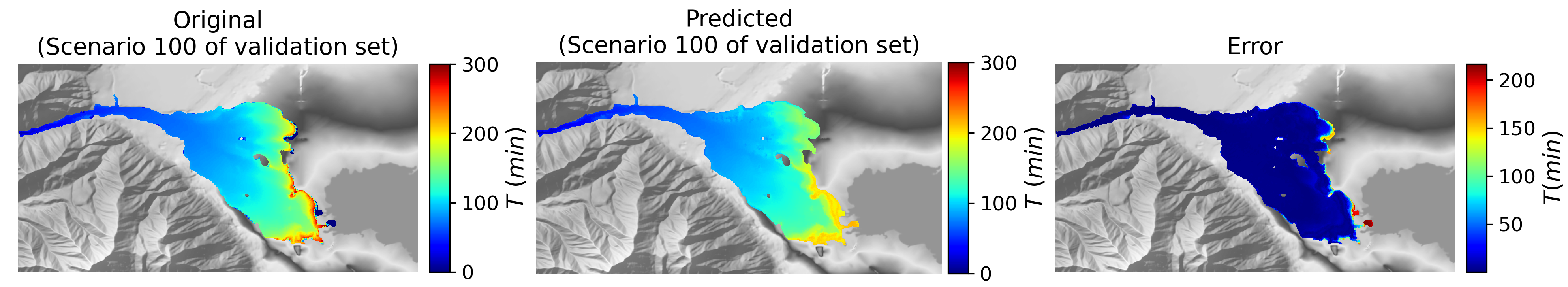}
    \caption{Comparison of filtered original validation map with filtered map obtained via GPR prediction and PCA-reconstruction.}
    \label{fig:app_prediction}
\end{figure}

Figure \ref{fig:app_prediction} shows an example of application of the aforementioned steps to predict an arrival times map. We note that the arrival time field is well recovered within the flooded region, as well as the wet-dry interface. The most relevant errors are near the interface.

\section{Validation of statistical supports and methods}
\label{app:validation}

\begin{figure}[htbp]
\centering

\subfloat[Yield stress $\yieldStress$.]{
    \includegraphics[width=0.7\textwidth]{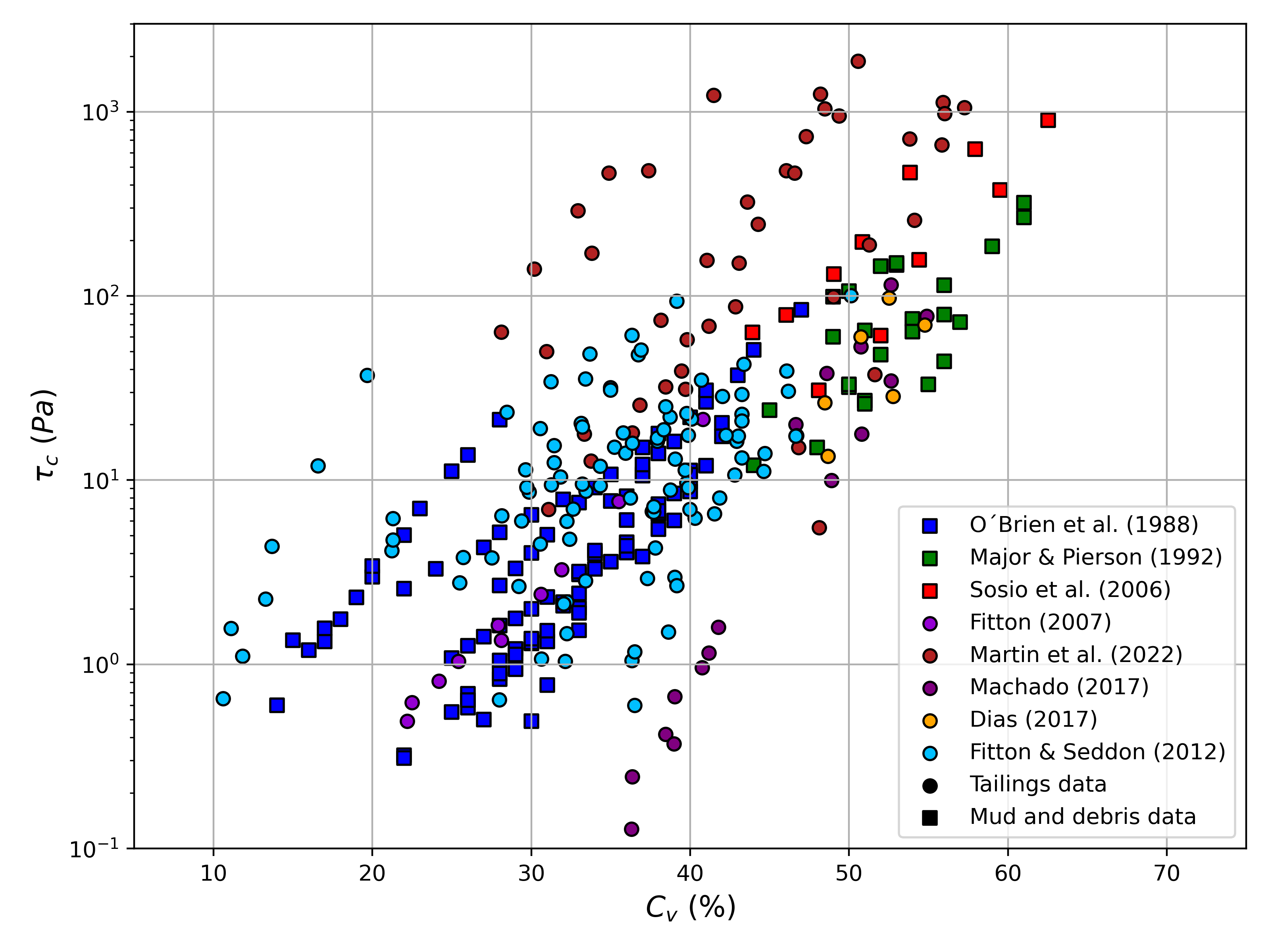}
    \label{fig:support_yield}
}
\hfill
\subfloat[Plastic viscosity $\plasticViscosity$.]{
    \includegraphics[width=0.7\textwidth]{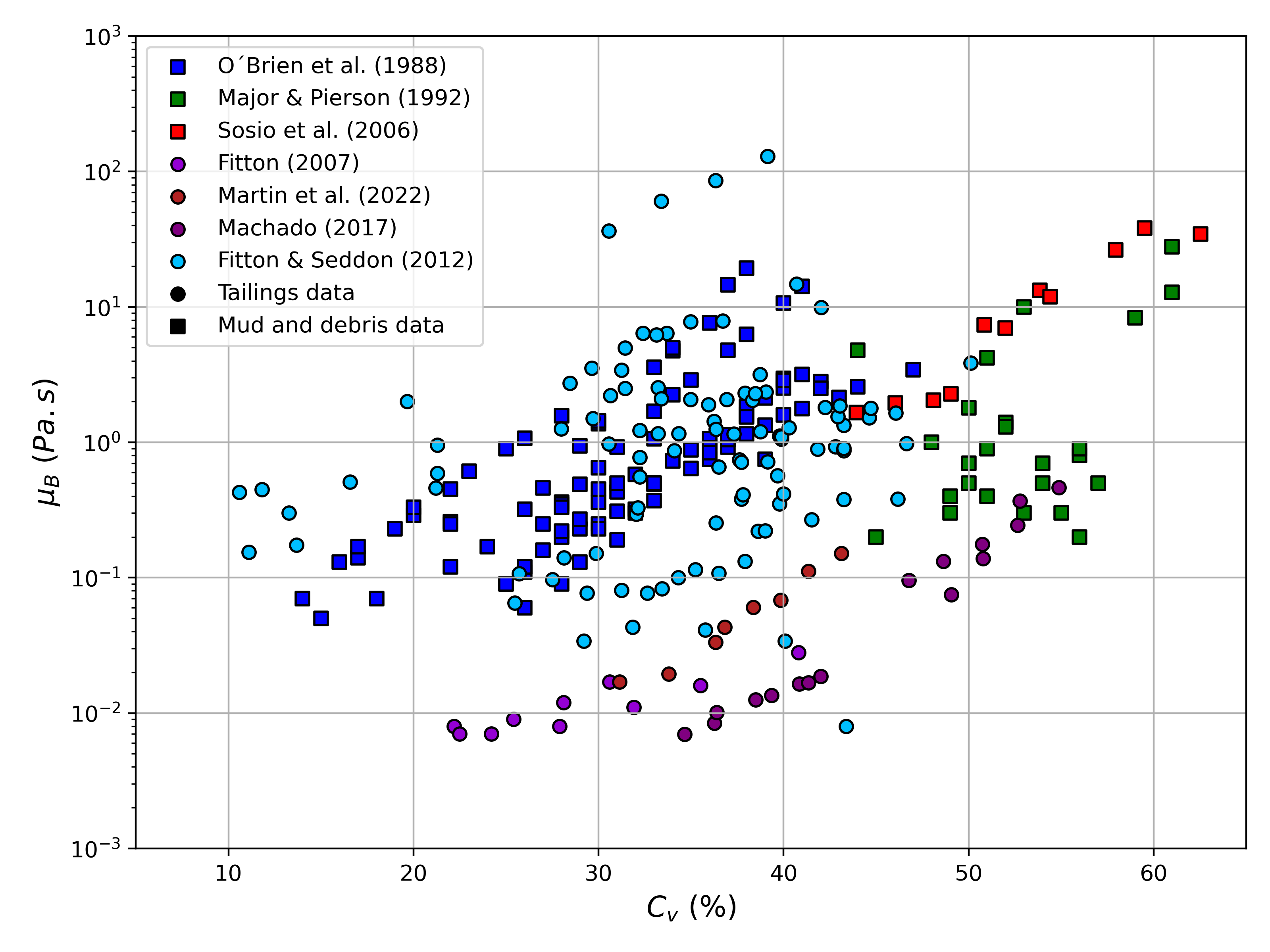}
    \label{fig:support_viscosity}
}

\caption{Rheological-parameter data compiled from the literature \citep{Obrien1988, Major1992, Sosio2006, Fitton2007, Martin2022, Machado2017, Dias2017, Fitton2012}.  Only data for tailings, muds and debris-flow materials characterised using the Bingham model and conventional rheometry were included.}
\end{figure}

This appendix presents results that support the choice of parameters from the probabilistic framework: statistical support of tailings properties ($\yieldStress, \plasticViscosity, \volumetricConc$), size of training and validation sets, number of basis components and size of PF samples. Concerning the statistical support of tailings properties, Figs.~\ref{fig:support_yield} and \ref{fig:support_viscosity} compile literature data of tailings and similar materials for $\yieldStress$ and $\plasticViscosity$, respectively. Care was taken to select data with a coherent procedure. The nature of the material was divided in mud/debris materials and tailings materials; only fine-grained materials were selected; only classical rheometry protocols were considered (results of back-analysis and of alternative rheometry techniques were discarded); and the adjusted rheological model was the Bingham model.

\begin{figure}[htbp]
\centering

\subfloat[First-order GSI.]{
    \includegraphics[width=0.7\textwidth]{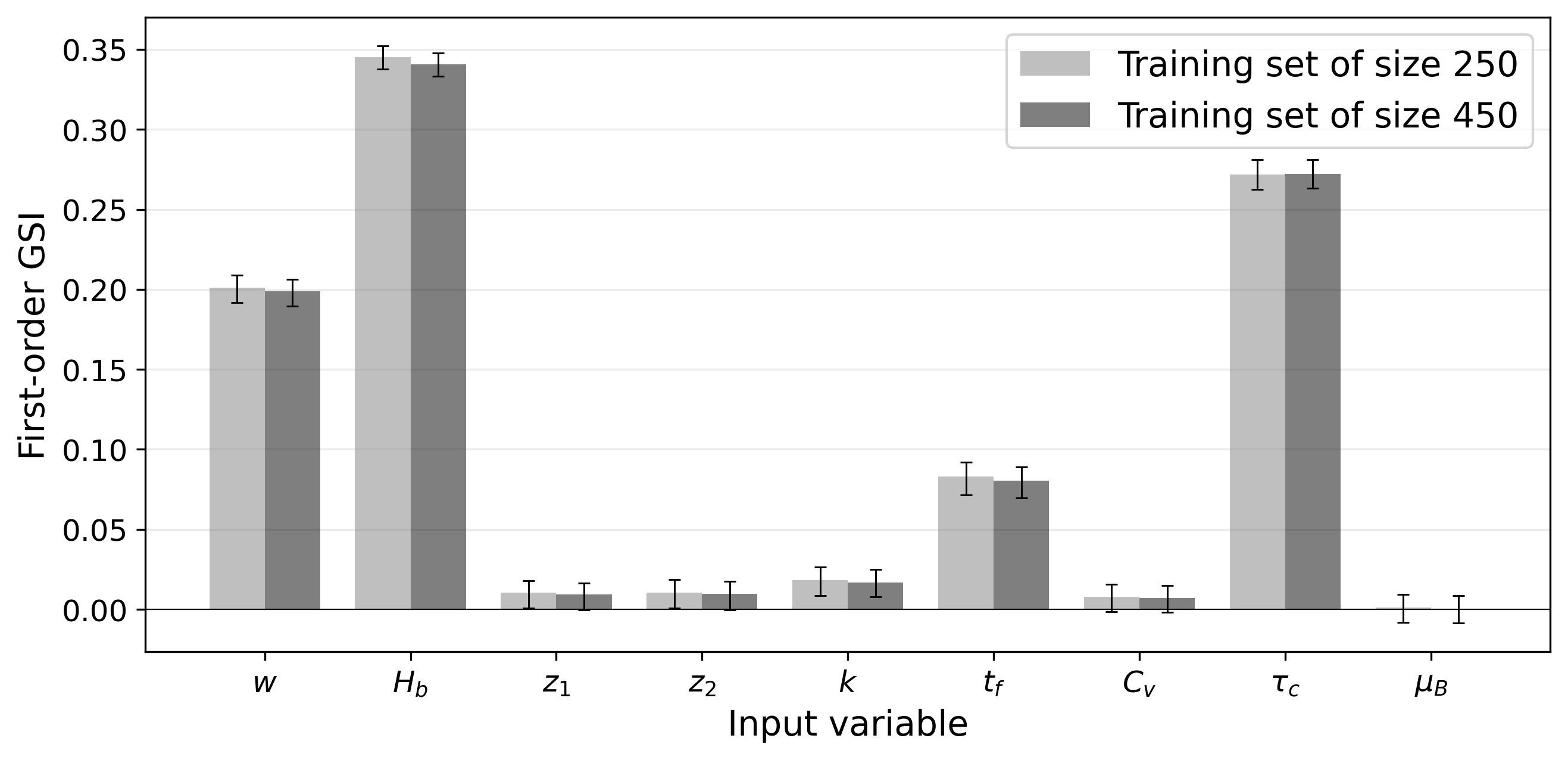}
    \label{fig:GSI_first_order_250_450}
}
\hfill
\subfloat[Total GSI.]{
    \includegraphics[width=0.7\textwidth]{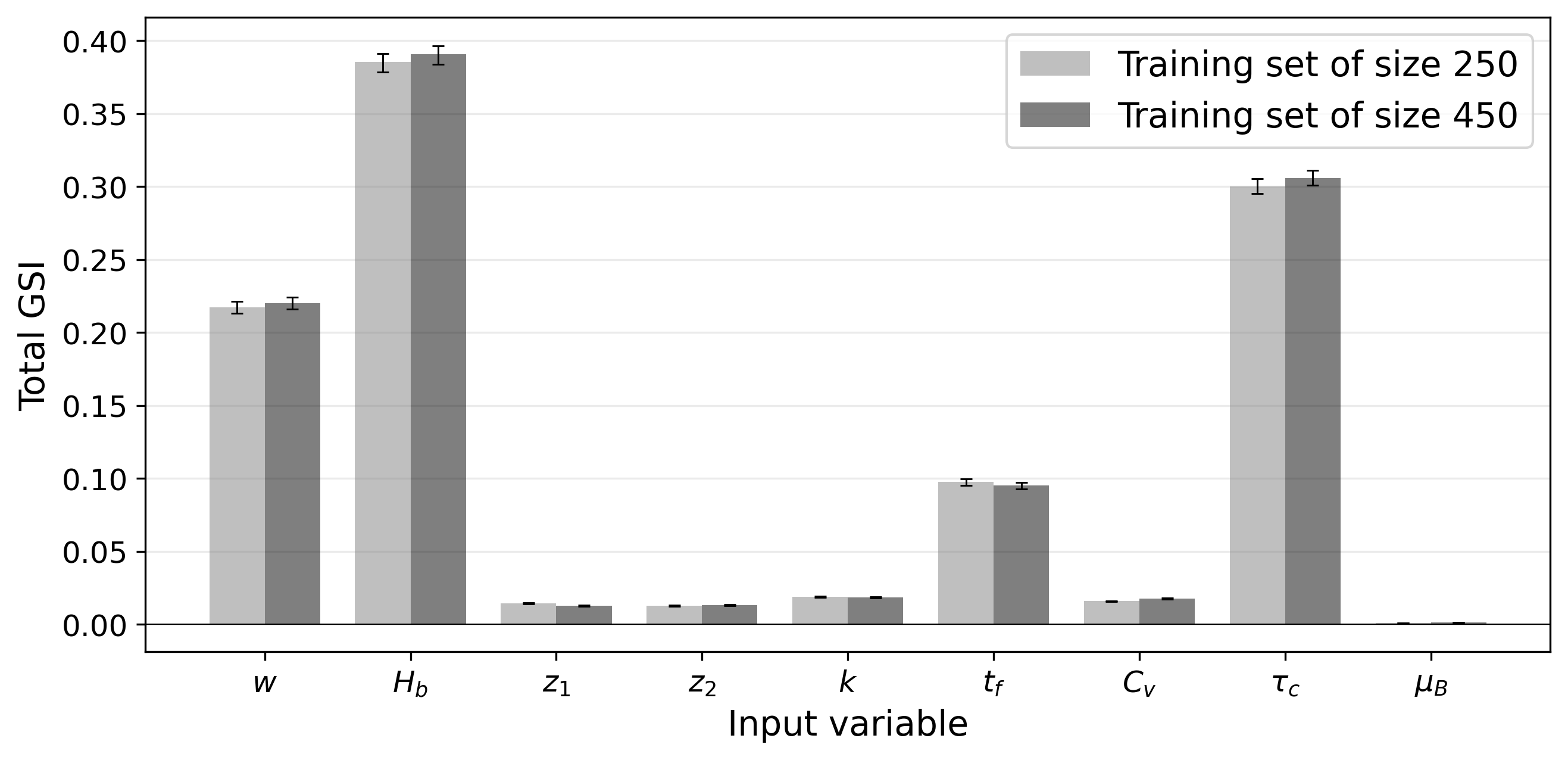}
    \label{fig:GSI_total_250_450}
}

\caption{Comparison between two sizes of training sets ($250$ and $450$) for first-order and total GSI.}
\label{fig:GSI_comparison}

\end{figure}

Regarding the training set, Fig. \ref{fig:GSI_comparison} shows the comparison between sets of sizes $250$ and $450$ for first-order and total GSI of maximum flow depth. No significant differences were found by increasing the training set to $450$; thus, a training set of size $250$ was chosen.

\begin{figure}[htbp]
\centering

\subfloat[Maximum flow depth.]{
    \includegraphics[width=0.45\textwidth]{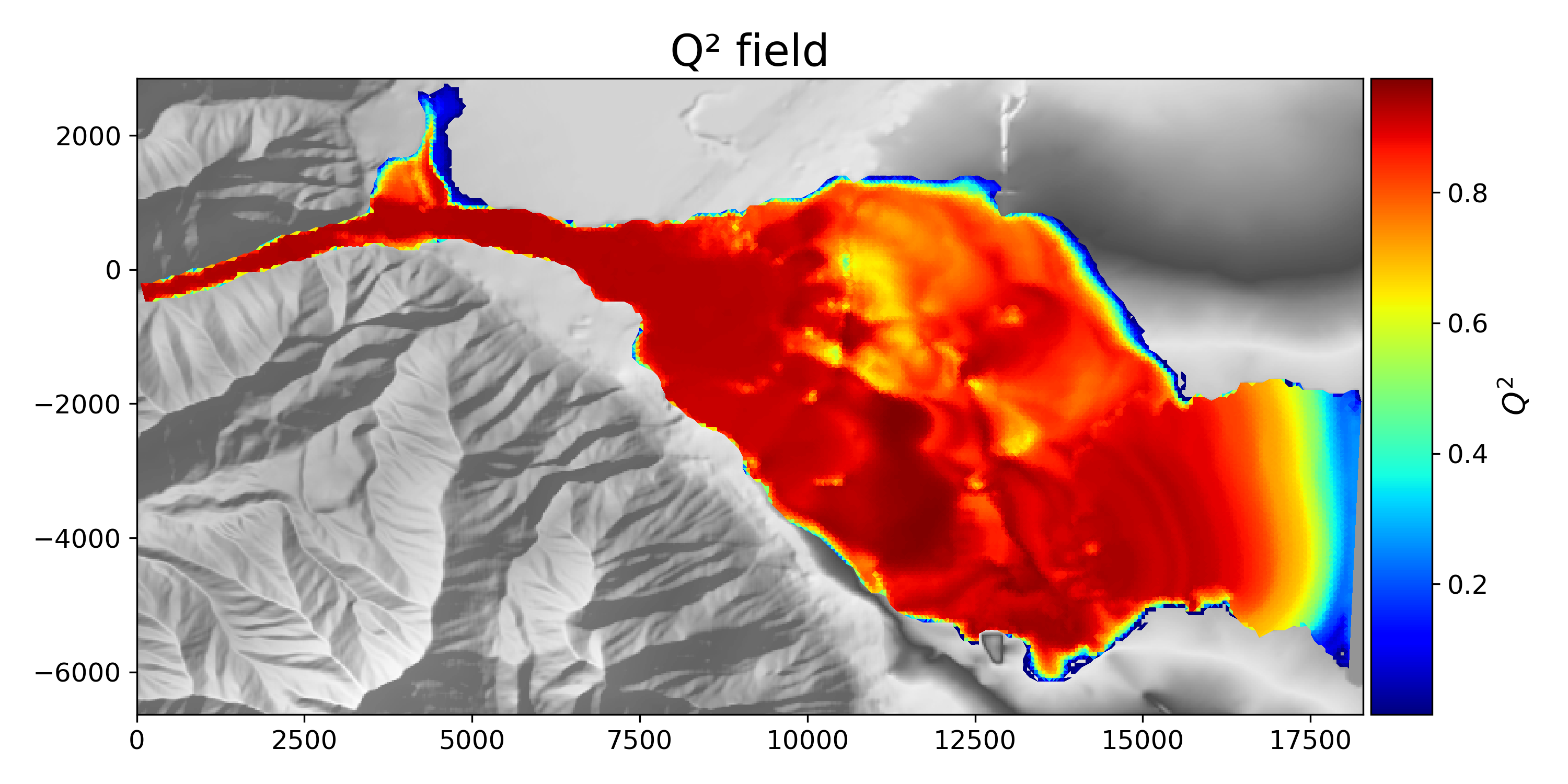}
    \label{fig:q2_max_depth}
}
\hfill
\subfloat[Arrival times.]{
    \includegraphics[width=0.45\textwidth]{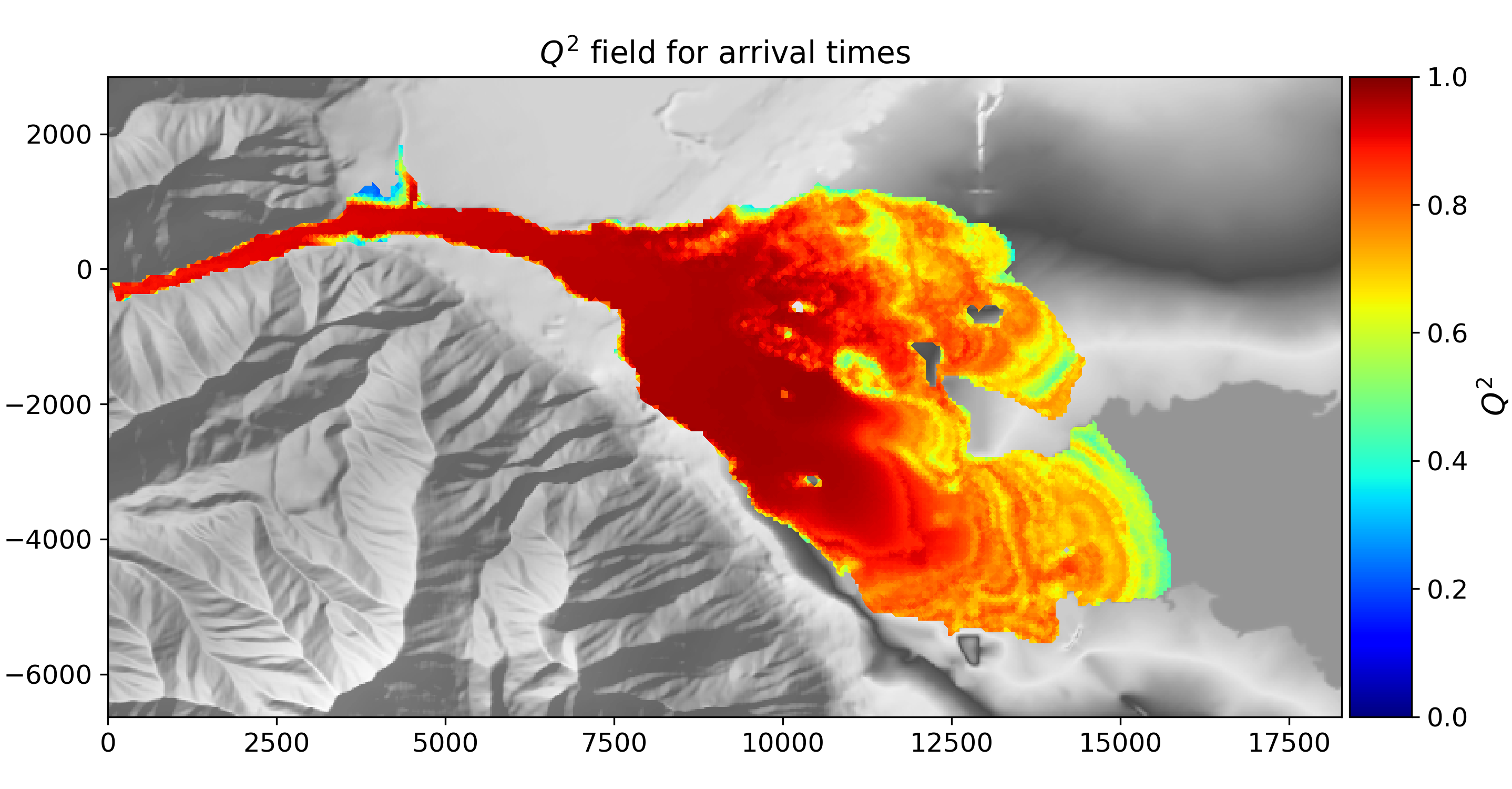}
    \label{fig:q2_arrival}
}

\caption{Spatially distributed $Q^2$ coefficients. Warm colours indicate better agreement between metamodel predictions and validation data.}
\label{fig:Q2}

\end{figure}

Concerning validation of metamodels, the set of metamodels trained with a training sample of $\nDoE=250$ are used to reconstruct maps through Eq. \ref{Eq:PCA_reconstruction_map} and compared with a validation dataset of size $250$. The performance of the set of metamodels was quantified by using the Nash-Sutcliffe coefficient ($Q^2$ metric) \citep{Marrel2011}, where a validation set of size $250$ was used. Figure \ref{fig:Q2} shows the spatial distribution of $Q^2$ for both maximum flow depth and arrival time maps. The $Q^2$ coefficient ranges from $-\infty$ to $1$. If $Q^2 > 0$, the prediction is better than the mean value of the validation set. The closer $Q^2$ is to $1$, the better the accuracy of the prediction. No significant difference was verified by using different training and validation scenarios to compute the $Q^2$ distribution. For the maximum flow depth, we observe elevated $Q^2$ for most cells and, therefore, good predictions globally. Downstream and boundary cells tend to show lower $Q^2$ since fewer scenarios are likely to reach them, making these cells harder to predict by metamodels. In the case of arrival times, the number of validation points of each cell is not the same due to the highly variable wet-dry interface. Thus, the $Q^2$ map was defined within cells with a minimum count of validation points of $50$. Most of the domain presents acceptable $Q^2$ scores, even though some parts of the domain are not defined.

\begin{figure}[htbp]
\centering

\subfloat[Maximum flow depth.]{
    \includegraphics[width=0.6\textwidth]{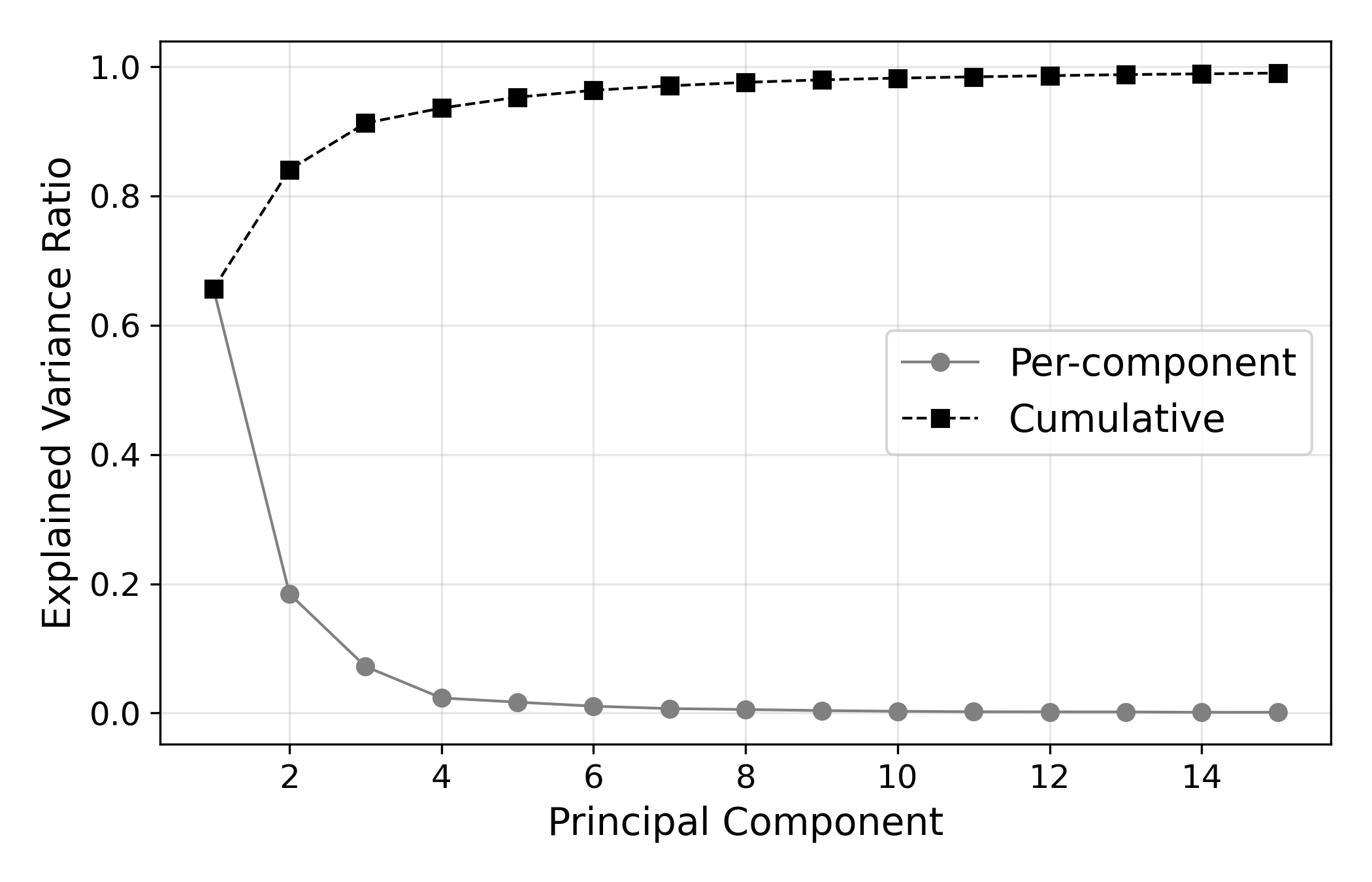}
    \label{fig:max_depth}
}
\hfill
\subfloat[Mean-imputed arrival times.]{
    \includegraphics[width=0.6\textwidth]{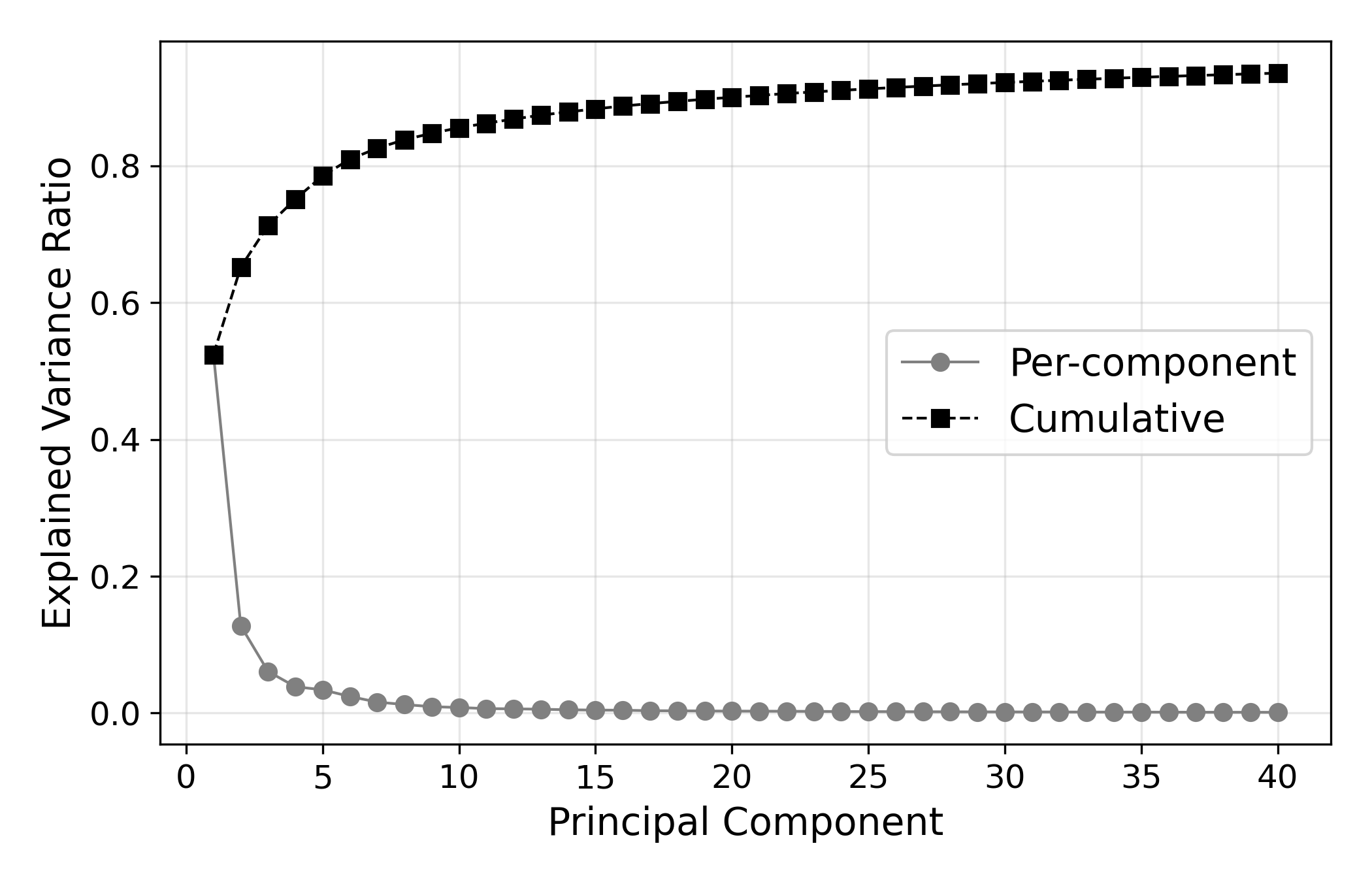}
    \label{fig:arrival_time}
}

\caption{Explained-variance ratio of each PCA component and the corresponding cumulative explained variance.}
\label{fig:explained_variance_PCA}

\end{figure}

Figure \ref{fig:explained_variance_PCA} shows typical explained variance of each PCA basis component, by considering the datasets of maximum flow depth and of the arrival times. For maximum flow depth, $\nbasis = 15$ is enough to represent more than $99 \%$ of the variance of output data. In the case of arrival time maps, we note that more components ($\nbasis = 40$) are needed to explain approximately $94 \%$ of output variance. Since arrival time maps have inherent issues, such as sharp wet-dry interfaces (see Appendix \ref{appendix_arrival_times}), more components are needed to represent these maps. However, this choice is not prohibitive for the methodology, since it is still a low-dimensionality problem.

\begin{figure}[H]
\centering

\subfloat[Maximum flow depth.]{
    \includegraphics[width=0.7\textwidth]{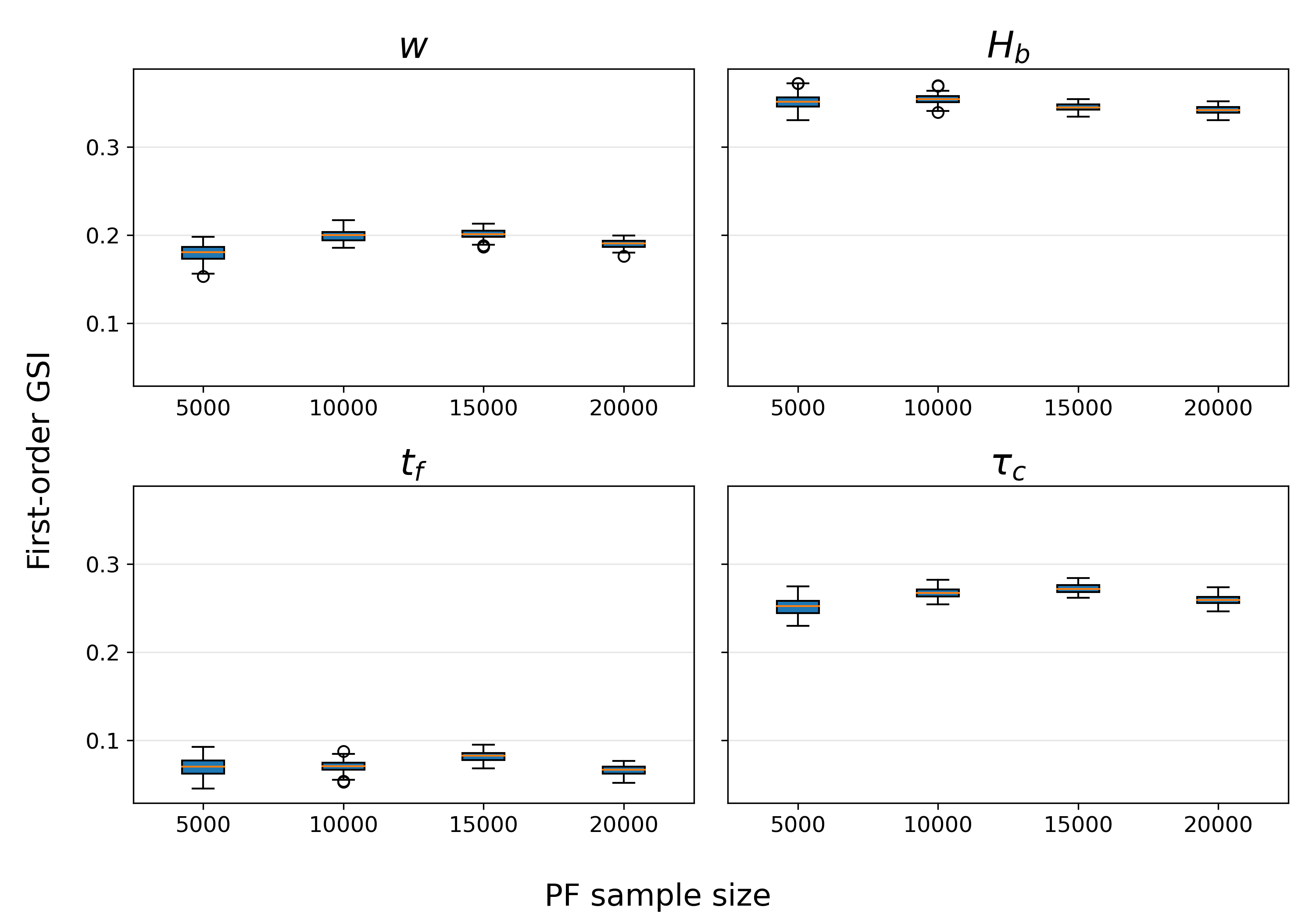}
    \label{fig:pf_gsi_max_depth}
}
\hfill
\subfloat[Arrival times.]{
    \includegraphics[width=0.7\textwidth]{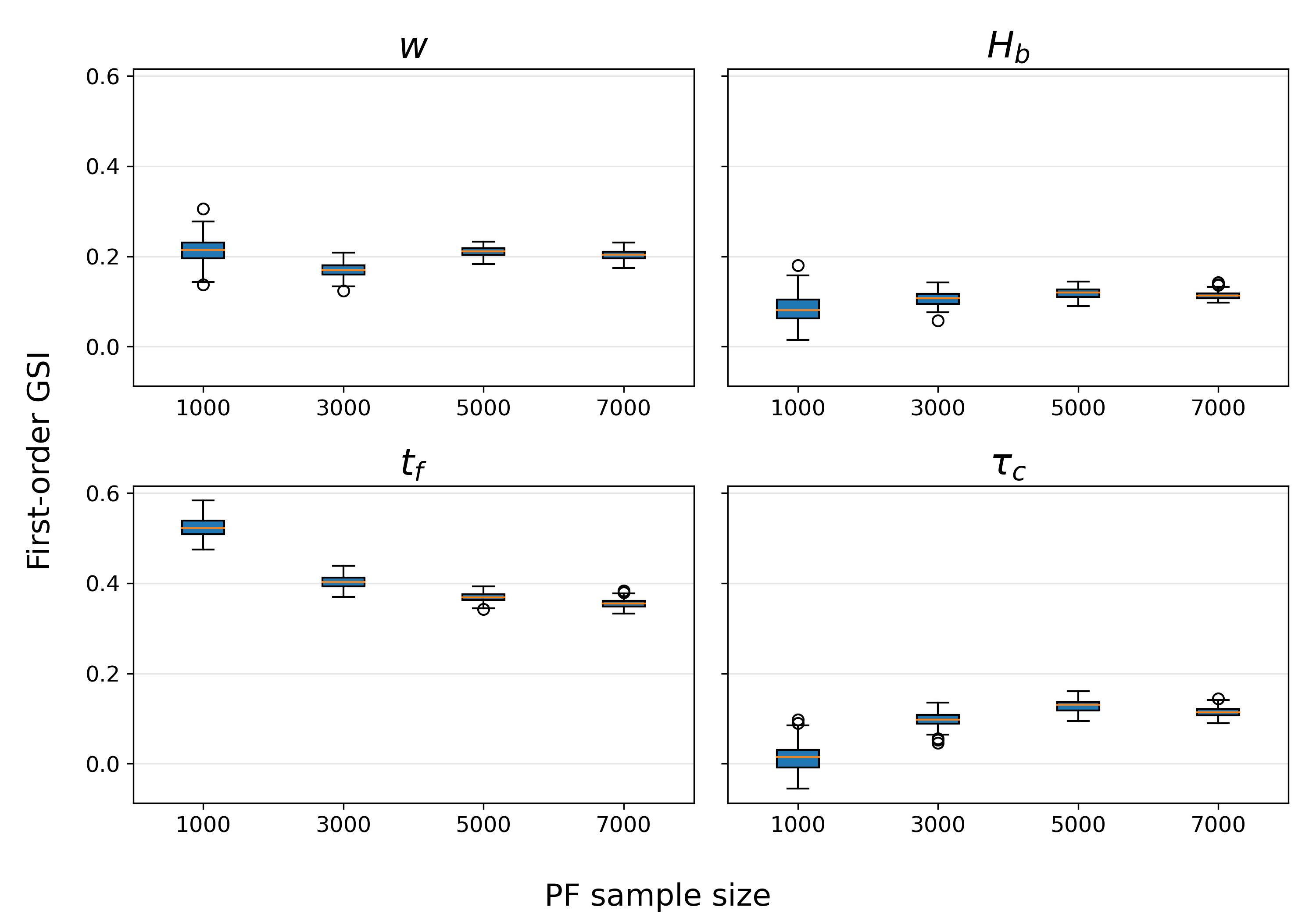}
    \label{fig:pf_gsi_arrival}
}

\caption{Sensitivity of the estimated first-order GSIs to the PF sample size.}
\label{fig:PF_verification}

\end{figure}

Concerning empirical estimation of sensitivity indices, Fig. \ref{fig:PF_verification} presents a verification of different sizes of PF samples using the first-order GSI as parameter. We note that the indices do not vary significantly as a function of the size, as long as $\nPF>5000$ for maximum flow depth maps and $\nPF>3000$ for arrival time maps. Therefore, the PF estimation performed in this work is considered to be stable over the tested PF sample-size range.

\printcredits

\section*{Declaration of competing interests}
The contact author has declared that none of the authors has any competing interests.

\section*{Data availability}
Data will be made available on request.

\section*{Acknowledgements}
The authors gratefully acknowledge Olivier Roustant (Institut National des Sciences Appliquées de Toulouse – INSA Toulouse, Department of Mathematics and Applications) for his teaching, guidance and insightful discussions on the probabilistic aspects of this work. This research was supported by the Fundação de Amparo à Pesquisa do Estado de São Paulo (FAPESP) through Ph.D. scholarship grants 2022/05184-1 and 2023/08472-0. 

\bibliographystyle{cas-model2-names}

\bibliography{cas-refs}



\end{document}